\documentclass[acmsmall,screen,authorversion=true, nonacm=true]{acmart}

\usepackage{xfrac}

\AtBeginDocument{%
  \providecommand\BibTeX{{%
    \normalfont B\kern-0.5em{\scshape i\kern-0.25em b}\kern-0.8em\TeX}}}

\setcopyright{none}
\acmDOI{}
\usepackage{subfig}
\usepackage{appendix}
\acmJournal{JACM}
\acmVolume{37}
\acmNumber{4}
\acmArticle{111}
\acmMonth{8}
\usepackage{listings}
\newcommand*{\doi}[1]{\href{https://doi.org/\detokenize{#1}}{doi: \detokenize{#1}}}
\begin{document}

\title{Time series compression survey}

\author{Giacomo~Chiarot}
\email{giacomo.chiarot@unive.it}
\orcid{1234-5678-9012}
\affiliation{%
  \institution{Department of Environmental Sciences, Informatics, and Statistics of Ca' Foscari University}
  \streetaddress{Via Torino 155}
  \city{Venice}
  \country{Italy}
  \postcode{30172}
}
\author{Claudio~Silvestri}
\orcid{0000-0002-9909-1808}
\authornotemark[1]
\email{silvestri@unive.it}
\affiliation{%
  \institution{Department of Environmental Sciences, Informatics, and Statistics of Ca' Foscari University}
  \streetaddress{Via Torino 155}
  \city{Venice}
  \country{Italy}
  \postcode{30172}
}
\affiliation{%
  \institution{European Centre for Living Technology (ECLT)}
  \streetaddress{Ca' Bottacin Dorsoduro 3911}
  \city{Venice}
  \country{Italy}
  \postcode{30123}
}

\renewcommand{\shortauthors}{Chiarot and Silvestri}

\begin{abstract}
    Smart objects are increasingly widespread and their ecosystem, also known as the Internet of Things, is relevant in many application scenarios. The huge amount of temporally annotated data produced by these smart devices demands efficient techniques for the transfer and storage of time series data.
    Compression techniques play an important role toward
    this goal and, even though standard compression methods could be used with some benefit, there exist several ones that specifically address the case of time series by exploiting their peculiarities to achieve more effective compression and more accurate decompression in the case of lossy compression techniques. 
    This paper provides a state-of-the-art survey of the principal time series compression techniques, proposing a taxonomy to classify them considering their overall approach and their characteristics. Furthermore, we analyze the performances of the selected algorithms by discussing and comparing the experimental results that were provided in the original articles. 
    
    The goal of this paper is to provide a comprehensive and homogeneous reconstruction of the state-of-the-art, which is currently fragmented across many papers that use different notations and where the proposed methods are not organized according to a classification.
\end{abstract}




\begin{center}
    \textcolor{red}{Published in: Giacomo Chiarot and Claudio Silvestri. 2022. Time series compression survey. ACM Comput. Surv. Just Accepted (August 2022). https://doi.org/10.1145/3560814} \\
\end{center} 

\maketitle

\fancyhead{} % clear all header fields
\fancyhead[CE, CO]{\textcolor{red}{Published in: Giacomo Chiarot and Claudio Silvestri. 2022. Time series compression survey. ACM Comput. Surv. Just Accepted (August 2022). https://doi.org/10.1145/3560814}}

\section{Introduction}\label{sec:introduction}
Time series are relevant in several contexts, and the Internet of Things ecosystem (IoT) is among the most pervasive ones. IoT devices, indeed, can be found in different applications, ranging from health care (smart wearables) to industrial ones (smart grids) \cite{asghari_internet_2019}, producing a large amount of time-series data. For instance, a single Boeing 787 fly can produce about half a terabyte of data from sensors \cite{ronkainen_designing_2015}. 
In those scenarios, characterized by high data rates and volumes, time series compression techniques are a sensible choice
to increase the efficiency of collection, storage, and analysis of such data. In particular, the need to include in the analysis both information related to the recent and the history of the data stream leads to consider data compression as a solution to optimize space without losing the most important information. A direct application of time series compression, for example, can be seen in Time Series Management Systems (or Time Series Database) in which compression is one of the most significant steps \cite{jensen_time_2017}.

There exists extensive literature on data compression algorithms, both on generic purpose ones for finite-size data and on domain-specific ones, for example, for images, video, and audio data streams. This survey aims at providing an overview of the state-of-the-art in time series compression research, specifically focusing on general-purpose data compression techniques that are either developed for time series or working well with time series. 

The algorithms we chose to summarize can deal with the continuous growth of time series over time and suitable for generic domains (as in the different applications in the IoT). Furthermore, these algorithms take advantage of the peculiarities of time series produced by sensors, such as:

\begin{itemize}
	\item \textbf{Redundancy}: some segments of a time series can frequently appear inside the same or other related time series;
	\item \textbf{Approximability}: sensors in some cases produce time series that can be approximated by functions;
	\item \textbf{Predictability}: some time series can be predictable, for example using deep neural network techniques.
\end{itemize}

\begin{figure}[ht!]
    \centering
    \includegraphics[width=300px]{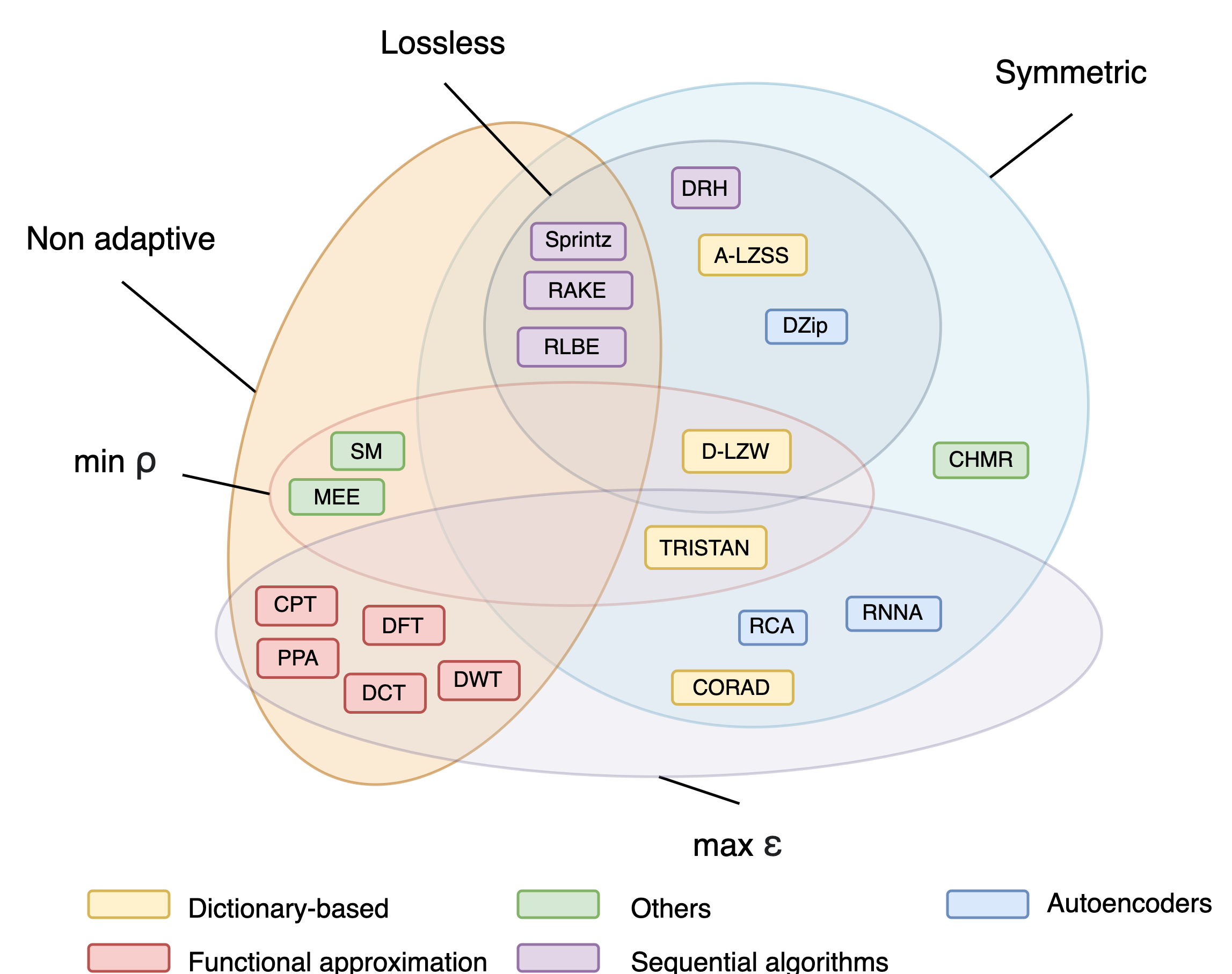}
    \caption{Visual classification of time series compression algorithms}
    \label{fig:venn}
\end{figure}

The main contribution of this survey is to present a reasoned summary of the state-of-the-art in time series compression algorithms, which is currently fragmented among several sub-domains ranging from databases to IoT sensor management.
Moreover, we propose a taxonomy of time series compression techniques based on their approach (dictionary-based, functional approximation, autoencoders, sequential, others) and their properties (adaptiveness, lossless reconstruction, symmetry, tuneability of max error or minimum compression ratio), anticipated in visual form in Figure~\ref{fig:venn} and discussed in Section~3, that will guide the description of the selected approaches. Finally, we recapitulate the results of performance measurements indicated in the described studies. 

\subsection{Outline of the Survey}

Section \ref{sec:literatureSelection} describes the method we applied to select the literature included in this survey. Section \ref{sec:background} provides some definitions regarding time series, compression, and quality indexes. Section \ref{sec:compressionTechniques} describes the compression algorithms and is structured according to the proposed taxonomy. Section~\ref{sec:experimentalResults} summarizes the experimental results found in the studies that originally presented the approaches we describe. A summary and conclusions are presented in Section~\ref{sec:Conclusions}.

\section{Literature selection} \label{sec:literatureSelection}

\subsection{Background and Motivation}

In the scientific literature, many surveys were published related to data compression, addressing particular domains (e.g., images, music, video) or type of compression (lossy or lossless), but there is not an exhaustive survey considering time series. 

\subsection{Literature Selection}

We conducted an extensive literature search on the ACM Digital Library using the query     \textit{[Title: serial sequences time series] AND [Title: compression]}, obtaining 2097 records. We carefully inspected the remaining papers and filtered out papers that were deemed to not be relevant. Furthermore, the survey also includes several additional papers hand-picked by the authors among the  relevant reference of other considered papers, by snowballing sampling, or published in the last 5 year in the same conference/journal as one of the considered paper.

\subsection{Scope}

In time series mining, time series analysis, and machine learning fields, there are many dimensionality reduction techniques that are used to achieve better results in classification and data analysis tasks \cite{krawczak_approach_2014, ali_concurrent_2021}. An overview of these techniques is presented in \cite{10.1145/882082.882086} with a detailed topology. Most of these techniques are not designed to reconstruct the original data after dimensionality reduction in a way that the error between the original and the reconstructed data is reasonably small, as, for example, natural language representation, trees, and random mapping. Our survey focuses on those techniques that allow to use the compressed representation to reconstruct the original data with a reasonably small or zero error.

\section{Background} \label{sec:background}

\subsection{Time series}
Time series are defined as a collection of data, sorted in ascending order according to the timestamp $t_i$ associated with each element. They are divided into:
\begin{itemize}
	\item Univariate Time Series (UTS): elements inside the collection are real values;
	\item Multivariate Time Series (MTS): elements inside the collections are arrays of real values, in which each position in the array is associated with a time series feature.
\end{itemize} 

\begin{figure}[ht]
	\centering
	\includegraphics[width=300px]{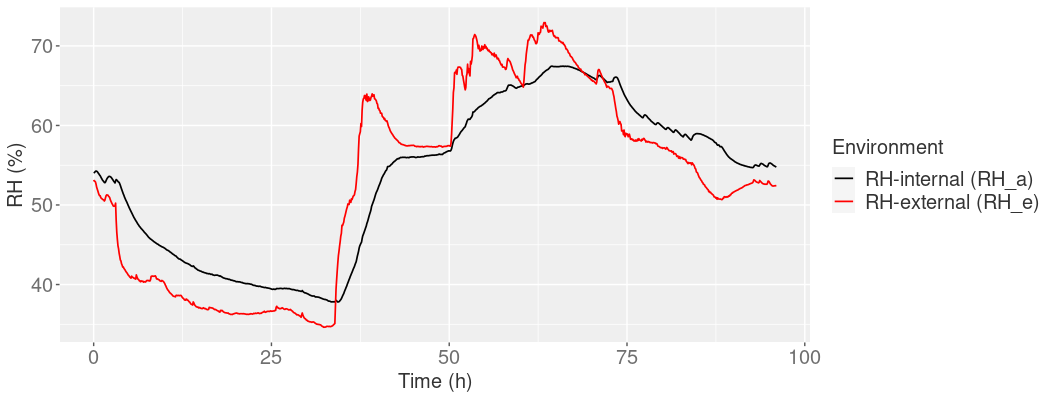}
	\caption{Example  of  an  MTS  representing  relative  humidity fluctuations inside and outside an artwork’s container}
	\label{fig:timeSeries}
\end{figure}

For instance, the temporal evolution of the average daily price of a commodity as the one represented in the plot in Figure~\ref{fig:timeSeries} can be modeled as a univariate time series, whereas the summaries of daily exchanges for a stock (including opening price, closing price, the volume of trades and other information) can be modeled as a multivariate time series.

Using a formal notation, time series can be written as:
\begin{equation}
	TS = [(t_1, x_1), \dots, (t_n, x_n)], x_i \in R^m
\end{equation}
where $n$ is the number of elements inside a time series and $m$ is the vector dimension of multivariate time series. For univariate time series, $m = 1$. Given $k\in [1, n]$, we write $TS[k]$ to indicate the $k$-th element $(t_n, x_n)$ of the time series $TS$.

A time series can be divided into segments, defined as a portion of the time series without any missing elements and ordering preserved: 
\begin{equation}
	TS_{[i,j]} = [(t_i, x_i), \dots, (t_j, x_j)]
\end{equation}
where $\forall k \in [i, j], TS[k] = TS_{[i,j]}[k-i+1]$.

\subsection{Compression}
\label{sub:compression}
Data compression, also known as \textit{source coding}, is defined in \cite{salomon_data_2007} as "the process of converting an input data stream (the source stream or the original raw data) into another data stream (the output, the bitstream, or the compressed stream) that has a smaller size". This process can take advantage of the \textit{Simplicity Power} (SP) theory, formulated in  \cite{wolff_computing_2003}, in which the compression goal is to remove redundancy while having high descriptive power. \par
The decompression process, complementary to the compression one,
%Complementary with compression is the decompression process. This process, 
is indicated also as \textit{source decoding}, and tries to reconstruct the original data stream from its compressed representation. \par
Compression algorithms can be described with the combination of different classes, shown in the following list:
\begin{itemize}
	\item \textbf{Non-adaptive - adaptive}: a non-adaptive algorithm that doesn't need a training phase to work efficiently with a particular dataset or domain since the operations and parameters are fixed, while an adaptive one does;
	\item \textbf{Lossy - lossless}: algorithms can be lossy if the decoder doesn't return a result that is identical to original data, or lossless if the decoder result is identical to original data;
	\item \textbf{Symmetric - non-symmetric}: an algorithm is symmetric if the decoder performs the same operations of the encoder in reverse order, whereas a non-symmetric one uses different operations to encode and decode a time series.
\end{itemize}
In the particular case of time series compression, a compression algorithm (\textit{encoder}) takes in input one Time Series $TS$ of size $s$ and returns its compressed representation $TS'$ of size $s'$, where $s' < s$ and the size is defined as the bits needed to store the time series: $E(TS) = TS'$. From the compressed representation $TS'$, using a \textit{decoder}, it is possible to reconstruct the original time series: $D(TS') = \overline{TS_s}$. If $\overline{TS} = TS_s$ then the algorithm is lossless, otherwise it is lossy.

In Section~\ref{sec:compressionTechniques}, there are shown the most relevant categories of compression techniques and their implementation.

\subsection{Quality indices}
To measure the performances of a compression encoder for time series, three characteristics are considered: compression ratio, speed, and accuracy.

\textbf{Compression ratio}
This metric measures the effectiveness of a compression technique, and it is defined as:
\begin{equation}
	\label{eq:compressionRatio}
	\rho = \frac{s'}{s}
\end{equation} 
where $s'$ is the size of the compressed representation and $s$ is the size of the original time series. Its inverse $\frac{1}{\rho}$ is named \textit{compression factor}. An index used for the same purpose is the \textit{compression gain}, defined as:
\begin{equation}
	c_g = 100\log_e \frac{1}{\rho}
\end{equation}

\textbf{Accuracy}, also called distortion, measures the fidelity of the reconstructed time series respect to the original. It is possible to use different metrics to determine fidelity~\cite{sayood_introduction_2006}:
\begin{itemize}
	\item \textbf{Mean Squared Error}: $MSE = \frac{\sum_{i = 1}^{n}{(x_i - \overline{x}_i)^2}}{n}$
	\item \textbf{Root Mean Squared Error}: $RMSE = \sqrt{MSE}$
	\item \textbf{Signal to Noise Ratio}: $SNR = \frac{\sum_{i = 1}^n{\sfrac{x_i^2}{n}}}{MSE}$
	\item \textbf{Peak Signal to Noise Ratio}: $PSNR = \frac{x_{pick}^2}{MSE}$ where $x_{peak}$ is the maximum value in the original time series.
\end{itemize}

\section{Compression algorithms} \label{sec:compressionTechniques} 

In this section, we present the most relevant time series compression algorithms by describing in short summaries their principles and peculiarities. We also provide a pseudo-code for each approach, focusing more on style homogeneity than on the faithful reproduction of the original pseudo-code proposed by the authors. For a more detailed description, we refer the reader to the complete details available in the original articles. Below the full list of algorithms described in this section, divided by approach:

\renewcommand{\labelenumii}{\theenumii}
\renewcommand{\theenumii}{\theenumi.\arabic{enumii}.}
\begin{enumerate}
    \item Dictionary-Based (DB):
    \begin{enumerate}
        \item TRISTAN;
        \item CORAD;
        \item A-LZSS;
        \item D-LZW.
    \end{enumerate}
    \item Functional Approximation (FA)
    \begin{enumerate}
        \item Piecewise Polynomial Approximation (PPA);
        \item Chebyshev Polynomial Transform (CPT);
        \item Discrete Wavelet Transform (DWT);
        \item Discrete Fourier Transform (DFT);
        \item Discrete Cosine Transform (DCT).
    \end{enumerate}
    \item Autoencoders:
    \begin{enumerate}
        \item Recurrent Neural Network Autoencoder (RNNA);
        \item Recurrent Convolutional Autoencoder (RCA);
        \item DZip.
    \end{enumerate}
    \item Sequential Algorithms (SA):
    \begin{enumerate}
        \item Delta encoding, Run-length and Huffman (DRH);
        \item Sprintz;
        \item Run-Length Binary Encoding (RLBE);
        \item RAKE.
    \end{enumerate}
    \item Others:
    \begin{enumerate}
        \item Major Extrema Extractor (MEE);
        \item Segment Merging (SM);
        \item Continuous Hidden Markov Chain (CHMC).
    \end{enumerate}
\end{enumerate}

RNNA algorithm can be applied both to univariate and multivariate time series. The other algorithms can also handle multivariate time series by extracting each feature as an independent time series. All the algorithms accept time series represented with real values.

\begin{figure}[ht]
	\centering
	\includegraphics[width=350px]{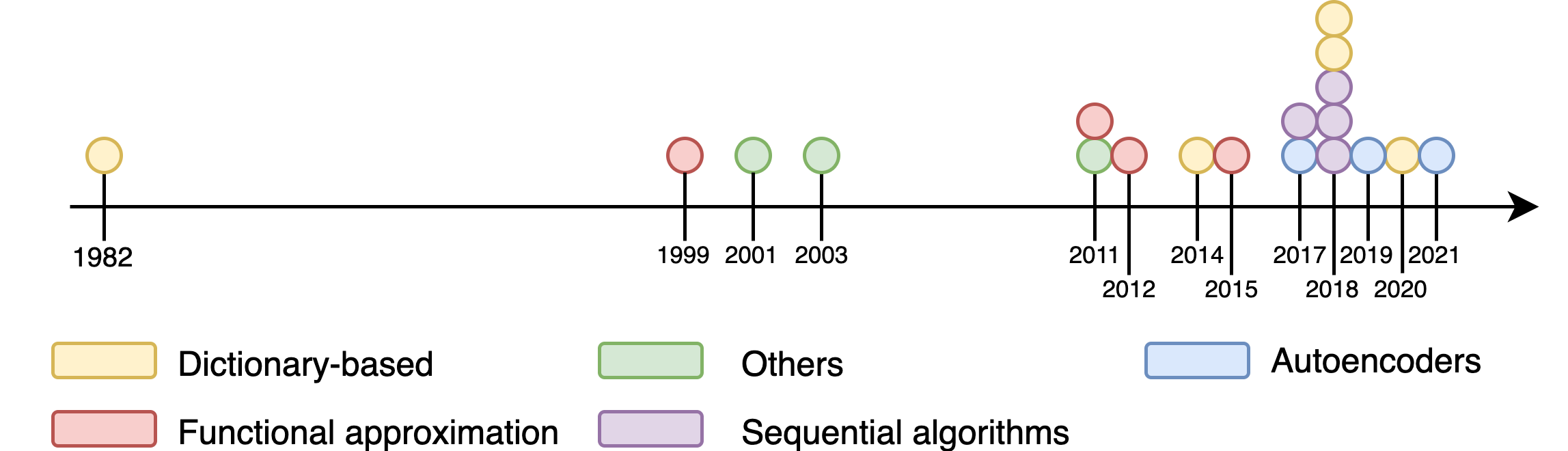}
	\caption{Chronological trend of compression techniques for time series}
	\label{fig:timeLine}
\end{figure}

The methods we considered span a temporal interval of more than 20 years, with a significant outlier dating back to the '80s. Figure \ref{fig:timeLine} graphically represents the temporal trend of adoption of the different approaches for the methods we considered. 

\subsection{Dictionary-Based (DB)} \label{subsec:dictionary-based}
This approach is based on the principle that time series share some common segments, without considering timestamps. These segments can be extracted into atoms, such that a time series segment can be represented with a  sequence of these atoms. Atoms are then collected into a dictionary that associates each atom with a univocal key used both in the representation of time series and to search efficiently their content. The choice of atoms length should guarantee a low decompression error and maximize the compression factor at the same time.
Algorithm~\ref{code:DB-training} shows how the training phase works at a high level: \texttt{createDictionary} function computes a dictionary of segments given a dataset composed by time series and a threshold value \texttt{th}.

The \texttt{find} function searches in the dictionary if there exists a segment which is similar to the segment \texttt{s} with a distance lower than threshold \texttt{th}; a possible index of distance can be $MSE$. If a match is found, the algorithm merges the segment \texttt{s} with the matched one to achieve generalization. Larger \texttt{th} value results in higher compression and lower reconstruction accuracy, and a dictionary with lower dimension. \par
After the segment dictionary is created, the compression phase takes in input one time series, and each segment is replaced with its key in the dictionary. If some segments are not present in the dictionary, they are left uncompressed, or a new entry is added to the dictionary. Algorithm~\ref{code:DB-compression} shows how the compression phase works: the \texttt{compress} function takes in input a time series to compress, the dictionary created during the training phase, a threshold value, and returns the compressed time series as a list of indices and segments.

The compression achieved by this technique can be either lossy or lossless, depending on the implementation. \par
The main challenges for this architecture are to:
\begin{itemize}
	\item maximize the searching speed to find time series segments in the dictionary;
	\item make the time series segments stored in the dictionary more general as possible to minimize the distance in the compression phase.
\end{itemize} \par

\subsubsection{TRISTAN}
One implementation of the Dictionary-Based architecture is TRISTAN
 \cite{marascu_tristan_2014}, an algorithm divided into two phases: the learning phase and the compression phase.

\textbf{Learning phase}
The dictionary used in this implementation can be created by domain experts that add typical patterns or by learning a training set. To learn a dictionary from a training set $T = [t_1, \dots, t_n]$ of $n$ segments, the following minimization problem has to be solved:
\begin{equation}
	\label{eq:TRISTAN-dictionary}
	\begin{split}
		D = \arg\underset{D}{\min}\sum_{i = 1}^{n}{\|w_i D - t_i\|_2} \\
		\text{Under the constraint: } \|w_i\|_0 \leq sp
	\end{split}
\end{equation}
where $D$ is the obtained dictionary, $sp$ is a fixed parameter representing sparsity, $w_i$ is the compressed representation of segment $i$, and $w_i D$ is the reconstructed segment. The meaning of this formulation is that the solution to be found is a dictionary that minimizes the distance between original and reconstructed segments. \par
The problem shown in equation~\ref{eq:TRISTAN-dictionary} is NP-hard \cite{marascu_tristan_2014}, thus an approximate result is computed. A technique for approximating this result is shown in \cite{mairal_online_2009}.

\textbf{Compression phase}
Once the dictionary is built, the compression phase consists in finding $w$ such that:
\begin{equation}
	\label{eq:TRISTAN-compression}
	s = w\cdot D
\end{equation}
where $D$ is the dictionary, $s$ is a segment and $w \in \{0,1\}^k$, and $k$ is the length of the compressed representation. An element of $w$ in position $i$ is $1$ if the $a_i \in D$ is used to reconstruct the original segment, $0$ otherwise. \par
Finding a solution for Equation~\ref{eq:TRISTAN-compression} is an NP-hard problem and, for this reason, the matching pursuit method \cite{mallat_matching_1993} is used to approximate the original problem:
\begin{equation}
	\label{eq:TRISTAN-compression-approx}
	\begin{split}
		\overline{s} = \underset{w}{\arg}\min\| wD - s\|_2 \\
		\text{Under the constraint: } \|\overline{s}\|_0\leq sp
	\end{split}
\end{equation}
where $sp$ is a fixed parameter representing sparsity.

\textbf{Reconstruction phase}
Having a dictionary $D$ and a compressed representation $w$ of a segment $s$, it is possible to compute $\overline{s}$ as:
\begin{equation}
	\overline{s} = wD
\end{equation}

\subsubsection{CORAD}
This implementation extends the idea presented in TRISTAN \cite{khelifati_corad_nodate}. The main difference is that it adds autocorrelation information to get better performances in terms of compression ratio and accuracy. 	\par
Correlation between two time series $TS^A$ and $TS^B$ is measured with the Pearson correlation coefficient:
\begin{equation}
	r = \frac{\sum_i^n(x_i-\overline{x})(y_i-\overline{y})}{\sqrt{\sum_i^n(x_i-\overline{x})^2}\sqrt{\sum_i^n(y_i - \overline{y})^2}}
\end{equation}
where $x_i$ is an element of $TS^A_n$, $y_i$ is an element of $TS^B_n$ and $\overline{x},\overline{y}$ are mean values of the corresponding time series. This coefficient can be applied also to segments that have different ranges of values and $r \in [-1, q]$ where 1 expresses the maximum linear correlation, -1 the maximum linear negative correlation, and 0 no linear correlation. \par
Time series are divided into segments and time windows are set. For each window, correlation is computed between each segment belonging to it, and results are stored in a correlation matrix $M\in  {R}^{n \times n}$ where $n$ is the segment number of elements.

\textbf{Compression phase}
During this phase, segments are sorted from the less correlated to the most correlated. To do this, a correlation matrix $M$ is used. The metric used to measure how much one segment is correlated with all the others is the absolute sum of the correlations, computed as the sum of each row of $M$. \par
Knowing correlation information, a dictionary is used only to represent segments that are not correlated with others, as in TRISTAN implementation. While the other segments are represented solely using correlation information.

\textbf{Reconstruction phase}
The reconstruction phase starts with the segment represented with dictionary atoms while the others are reconstructed looking at the segment to which they are correlated. \par
This process is very similar to the one proposed in TRISTAN, with the sole difference that 
segments represented in the dictionary are managed differently than those that are not.

\subsubsection{Accelerometer LZSS (A-LZSS)}
Accelerometer LZSS is an algorithm built on top of the LZSS algorithm \cite{storer_data_1982} for searching matches \cite{pope_accelerometer_2018}. A-LZSS algorithm uses Huffman codes, generated offline using frequency distributions. In particular, this technique considers blocks of size $s=1$ to compute frequencies and build the code: an element of the time series will be replaced by a variable number of bits. Moreover, larger blocks can be considered and, in general, having larger blocks gives better compression performances at the cost of larger Huffman code tables. \par 
The implementation of this technique is shown in Algorithm \ref{ls:ALZSS}, where: 

\begin{itemize}
    \item \texttt{minM}: the minimum match length, which is asserted to be $\texttt{minM} > 0$;
    \item \texttt{Ln}: determines the lookahead distance, as $2^{Ln}$;
    \item \texttt{Dn}: determines the dictionary atoms length, as $2^{Dn}$;
    \item \texttt{longestMatch}: is a function that returns the index $I$ of the found match and the length L of the match. If the length of the match is too small, then the Huffman code representation of \texttt{s} is sent as the compression representation, otherwise, the index and the length of the match are sent, and the next L elements are skipped.
\end{itemize} 

This implementation uses a brute-force approach, with complexity $O(2^{Dn}\cdot 2^{Ln})$ but it is possible to improve over it by using hashing techniques.

\subsubsection{Differential LZW (D-LZW)}
The core of this technique is the creation of a very large dictionary that grows over time: once the dictionary is created, if a buffer block is found inside the dictionary it is replaced by the corresponding index, otherwise, the new block is inserted in the dictionary as a new entry \cite{tuong_ly_le_lossless_2018}.\par 
Adding new blocks guarantees lossless compression, but has the drawback of having too large dictionaries. This makes the technique suitable only for particular scenarios (i.e. input streams composed by words/characters or when the values inside a block are quantized). \par 
Another drawback of this technique is how the dictionary is constructed: elements are simply appended to the dictionary to preserve the indexing of previous blocks. For a simple implementation of the dictionary, the complexity for each search is $O(n)$ where $n$ is the size of the dictionary. This complexity can be improved by using more efficient data structures.
\subsection{Function Approximation (FA)} \label{subsec:PPA} The main idea behind function approximation is that a time series can be represented as a function of time. Since finding a function that approximates the whole time series is infeasible due to the presence of new values that cannot be handled, the time series is divided into segments and for each of them, an approximating function is found. \par 
Exploring all the possible functions $f: T \rightarrow X$ is not feasible, thus implementations consider only one family of functions and try to find the parameters that better approximate each segment. This makes the compression lossy.

A point of strength is that it does not depend on the data domain, so no training phase is required since the regression algorithm considers only single segments in isolation.

\subsubsection{Piecewise Polynomial Approximation (PPA)}\label{subsubsec:ppa}
This technique divides a time series into several segments of fixed or variable length and tries to find the best polynomials that approximate segments.

Despite the compression is lossy, a maximum deviation from the original data can be fixed a priori to enforce a given reconstruction accuracy. \par 
The implementation of this algorithm is described in \cite{eichinger_time-series_2015} where the authors apply a greedy approach and three different online regression algorithms for approximating constant functions, straight lines, and polynomials. These online algorithms are:
\begin{itemize}
	\item the PMR-Midrange algorithm, that approximates using constant functions \cite{lazaridis_capturing_2003};
	\item the optimal approximation algorithm, described in \cite{dalari_approximations_2006}, that uses linear regression;
	\item the randomized algorithm presented in \cite{seidel_small_1991}, that approximates using polynomials.
\end{itemize} \par 
The algorithm used in \cite{eichinger_time-series_2015} for approximating a time series segment is explained in Algorithm~\ref{ls:PPA}.

This algorithm finds repeatedly the polynomial of degree between 0 and a fixed maximum that can approximate the longest segment within the threshold error, yielding the maximum local compression factor. After a prefix of the stream has been selected and compressed into a polynomial, the algorithm analyzes the following stream segment. A higher value of $\epsilon$ returns higher compression and lower reconstruction accuracy. The fixed maximum polynomial degree $\rho$ affects compression speed, accuracy, and compression ratio: higher values slow down compression and reduce the compression factor, but return higher reconstruction accuracy.

\subsubsection{Chebyshev Polynomial Transform (CPT)}
Another implementation of polynomial compression can be found in \cite{hawkins_algorithm_2012}. In this article, the authors show how a time series can be compressed in a sequence of finite Chebyshev polynomials. \par 
The principle is very similar to the one shown in subsection~\ref{subsubsec:ppa} but based on the use of a different type of polynomial. Chebyshev Polynomials are of two types, $T_n(x)$, $U_n(x)$, defined as \cite{lv_chebyshev_2017}:
\begin{align}
	T_n(x) = \frac{n}{2}\sum_{k=0}^{n / 2}(-1)^k\frac{(n-k-1)!}{k!(n-2k)!}(2x)^{n-2k}, |x|<1 \\
	U_n(x) = \sum_{k=0}^{n/2}(-1)^k\frac{(n-k)!}{k!(n-2k)!}(2x)^{n-2k}, |x|<1
\end{align}
where $n \geq 0$ is the polynomial degree.

\subsubsection{Discrete Wavelet Transform (DWT)}
Discrete wavelet transform uses wavelet functions to transform time series. Wavelets are functions that, similarly to a wave, start from zero, and end with zero after some oscillation. An application of this technique can be found in \cite{olkkonen_ecg_2011}. \par 
This transformation can be written as:
\begin{equation}
	\Psi_{m,n}(t) = \frac{1}{\sqrt{a^m}}\Psi\left(\frac{t - nb}{a^m}\right)
\end{equation}
where $a > 1$, $b > 0$, $m,n\in {Z}$. \par 
To recover one transformed signal, the following formula can be applied:
\begin{equation}
	x(t) = \sum_{m\in  {Z}}\sum_{n\in  {Z}}\langle x, \Psi_{m,n}\rangle\cdot \Psi_{m,n}(t)
\end{equation} \par 

\subsubsection{Discrete Fourier Transform (DFT)}

Together with the Discrete wavelet transform, the Discrete Fourier transform is commonly used for signals compression. Given a time series $[(t_1, x_1), \dots, (t_n ,x_n)]$, it is defined as:

\begin{equation*}
    X_k = \sum^{N}_{n=1} x_n e^{\frac{-i2\pi kn}{N}}
\end{equation*}

Where $e^{\frac{i2\pi}{N}}$ is a primitive $N$th root of 1.

To efficiently compute this transformation, the Fast Fourier transform algorithm can be used, as the one introduced here \cite{cooley_algorithm_nodate}. As described in \cite{karim_wavelet_2011}, compressing a time series can be done after applying the DFT by cutting coefficients that are close to zero. This can be done also by fixing a compression ratio and discarding all the coefficients after a certain index.

\subsubsection{Discrete Cosine Transform (DCT)}

Given a time series $[(t_1, x_1), \dots, (t_n ,x_n)]$, the Discrete Cosine Transform is defined as:

\begin{equation*}
    c_k = \sum^{N}_{n=0} x_n \cdot \text{cos} \left[\frac{\pi}{N}\left(n + \frac 1 2\right)k\right]
\end{equation*}

The computation of this transformation is computationally expensive, so efficient implementations can be used, as the one presented in \cite{huang_refined_1999}.

Similarly to the DFT, compression of time series can be achieved by cutting the coefficients that are close to zero or by fixing a compression ratio and discarding all the coefficients after a certain index. An example of this technique can be found at \cite{mess_compression_nodate}.

\subsection{Autoencoders} \label{subsec:RNNA}
\begin{figure}[ht!]
    \centering
    \includegraphics[width=218px]{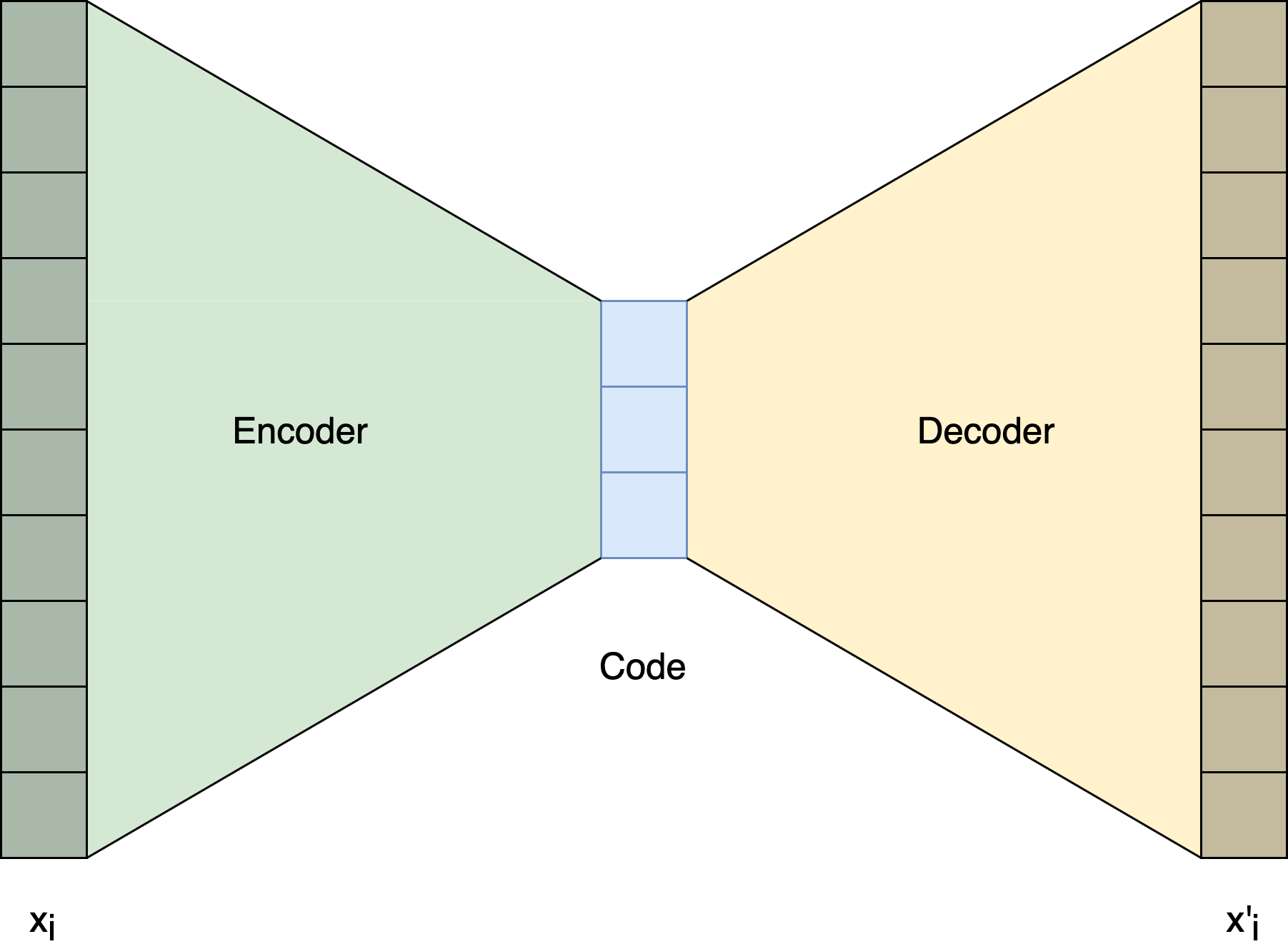}
    \caption{Simple autoencoder architecture}
    \label{fig:autoencoder}
\end{figure}

An autoencoder is a particular neural network that is trained to give as output the same values passed as input. Its architecture is composed of two symmetric parts: encoder and decoder. Giving an input of dimension $n$, the encoder gives as output a vector with dimensionality $m < n$, called code, while the decoder reconstructs the original input from the code, as shown in Figure~\ref{fig:autoencoder} \cite{Goodfellow-et-al-2016}. 

\subsubsection{Recurrent Neural Network Autoencoder (RNNA)}
\begin{figure}[ht!]
	\centering
	\includegraphics[width=218px]{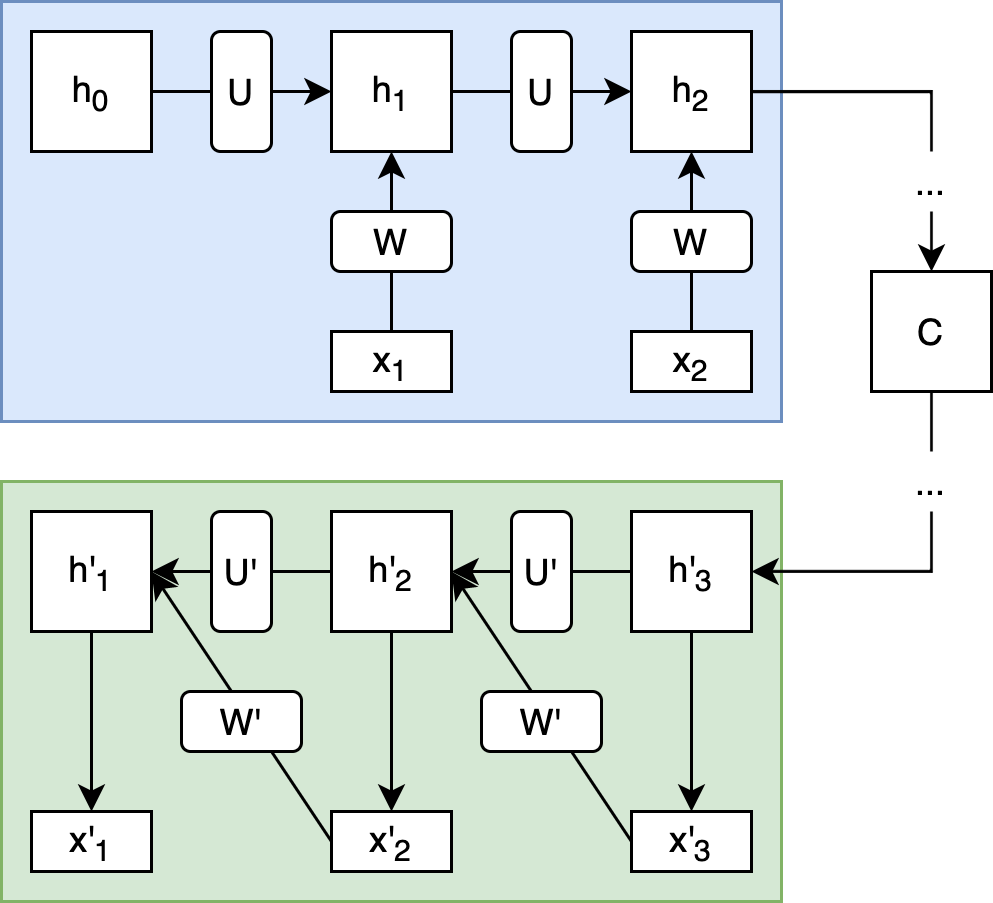}
	\caption{General structure of a RNN encoder and decoder}
	\label{fig:rnn}
\end{figure}

RNNA compression algorithms exploit recurrent neural networks \cite{sherstinsky_fundamentals_2020} to achieve a compressed representation of a time series. Figure~\ref{fig:rnn} shows the general unrolled structure of a recurrent neural network encoder and decoder. The encoder takes in input time series elements, which are combined with hidden states. Each hidden state is then computed starting from the new input and the previous state. The last hidden state of the encoder is passed as the first hidden state of the decoder, which applies the same mechanism, with the only difference that each hidden state provides an output. The output provided by each state is the reconstruction of the relative time series element and is passed to the next state.
 \par 
New hidden state $h_t$ is obtained by applying:
\begin{equation}
	h_t = \phi(Wx_t+Uh_{t-1})
\end{equation}
where $\phi$ is a logistic sigmoid function or the hyperbolic tangent. \par
One application of this technique is shown in \cite{hsu_time_2017}, in which also Long Short-Term Memory \cite{sepp_long_1997} is considered. This implementation compresses time series segments of different lengths using autoencoders. The compression achieved is lossy and a maximal loss threshold $\epsilon$ can be enforced. \par 
The training set is preprocessed considering temporal variations of data  applying:
\begin{equation}
	\label{eq:variance}
	\Delta(\mathcal{L}) = \sum_{t \in \mathcal{L}}|x_t-x_{t-1}|
\end{equation}
where $\mathcal{L}$ is the local time window. \par 
The value obtained by Equation~\ref{eq:variance} is then used to partition the time series, such that each segment has a total variation close to a predetermined value $\tau$. \par 
Algorithm~\ref{code:RRACA} shows an implementation of the RNN autoencoder approach, with an error threshold $\epsilon$, where:
\begin{itemize}
	\item \texttt{RAE a} is the recurrent autoencoder trained on a training set, composed of an encoder and a decoder;
	\item \texttt{getError} computes the reconstruction error between the original and reconstructed segment;
	\item $\epsilon$ is the error threshold value.
\end{itemize}

\subsubsection{Recurrent Convolutional Autoencoder (RCA)}

A variation of the technique presented in the previous subsection is proposed in \cite{zhao_time_2019} and consists of adding convolutional layers. Convolutional layers are used mainly for images data to extract local features. In the case of time series, this layer allows extracting local fluctuations, addressing complex inputs.

\subsubsection{DZip}

\begin{figure}[ht!]
	\centering
	\includegraphics[width=\columnwidth]{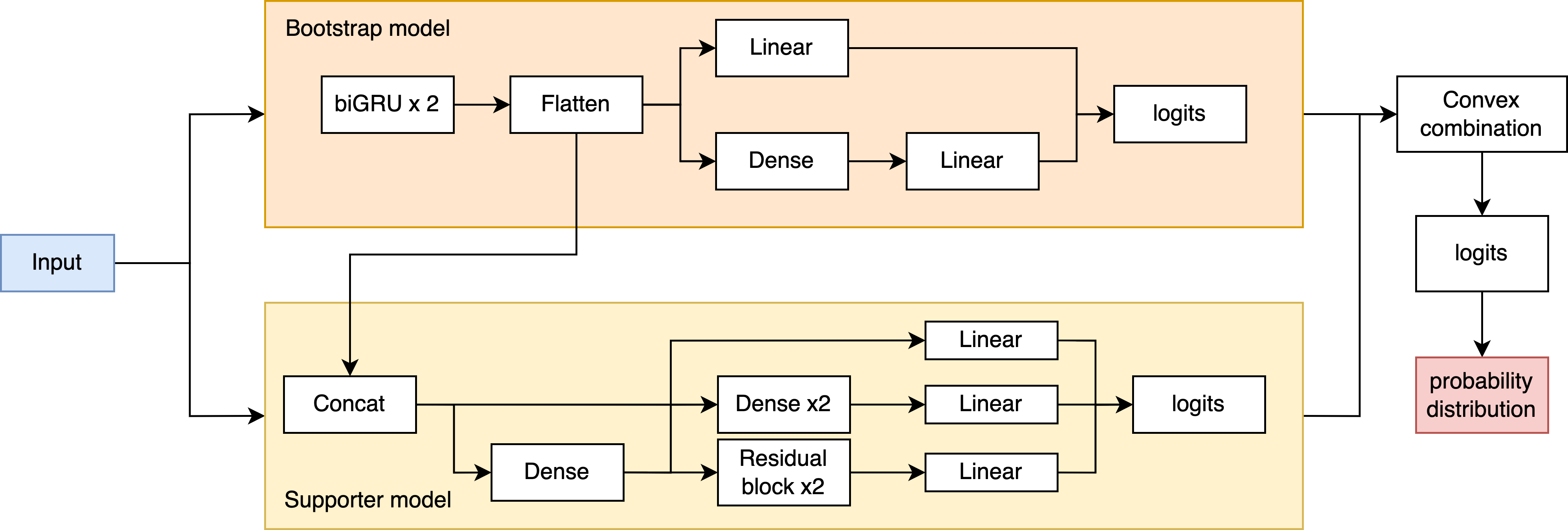}
	\caption{DZip model infrastructure}
	\label{fig:dzip}
\end{figure}

DZip is a lossless recurrent autoencoder \cite{goyal_dzip_2021}. This model also uses the prediction technique: it tries to predict the next symbol, having, in this case, a fixed vocabulary. To achieve a lossless compression, it combines two modules: the bootstrap model and the supporter model, as shown in Figure \ref{fig:dzip}.

 \par 

The bootstrap model is composed by stacking two bidirectional gated recurrent units, followed by linear and dense layers. The output of this module is a probability distribution, that is, the probability of each symbol in the fixed vocabulary to be the next one.

The supporter model is used to better estimate the probability distribution respect to the bootstrap model. It is composed by stacking three neural networks which act as independent predictors of varying complexity.

The two models are then combined by applying the following equation:

\begin{equation*}
    o = \lambda o_{b} + (1 - \lambda) o_s
\end{equation*}

Where $o_b$ is the output of the bootstrap model, $o_s$ is the output of the supporter model, and $\lambda \in [0, 1]$ is a learnable parameter.

\subsection{Sequential algorithms}\label{sec:sequentals}
This architecture is characterized by combining sequentially several simple compression techniques. Some of the most used techniques are:
\begin{itemize}
    \item Huffman coding;
    \item Delta encoding;
    \item Run-length encoding;
    \item Fibonacci binary encoding.
\end{itemize}
These techniques, summarized below, are the building blocks of the methods presented in the following subsections.

\textbf{Huffman coding}
\label{sub:HC}
Huffman coding is the basis of many compression techniques since it is one of the necessary steps, as for the algorithm shown in Subsection~\ref{subsub:rake}. \par 
The encoder creates a dictionary that associates each symbol with a binary representation and replaces each symbol of the original data with the corresponding representation. The compression algorithm is shown in Algorithm~\ref{code:HMCode} \cite{sayood_introduction_2006}.

The \texttt{createPriorityList} function creates a list of elements ordered from the less frequent to the most frequent, \texttt{addTree} function adds to the tree a father node with its children and \texttt{createDictionary} assigns 0 and 1 respectively to the left and right arcs and creates the dictionary assigning to characters the sequence of 0$|$1 in the route from the root to a leaf that represents the characters. Since the priority list is sorted according to frequency, more frequent characters are inserted closer to the root and they are represented using a shorter code.\par 
The decoder algorithm is very simple since the decoding process is the inverse of the encoding process: the encoded bits are searched in the dictionary and replaced with the corresponding original symbols.

\textbf{Delta encoding}
This technique encodes a target file with respect to one or more reference files \cite{sakr_delta_2018}. In the particular case of time series, each element at time $t$ is encoded as $\Delta(x_t, x_{t-1})$.

\textbf{Run-length}
In this technique, each run (a sequence in which the same value is repeated consecutively) is substituted with the pair $(v_t, o)$ where $v_t$ is the value at time $t$ and $o$ is the number of consecutive occurrences \cite{hardi_comparative_2019}.

\textbf{Fibonacci binary encoding}
This encoding technique is based on the Fibonacci sequence, defined as:
\begin{equation}
    F(N) = 
    \begin{cases} 
    F(N-1) + F(N-2), & \mbox{if } N >= 1  \\
    N , & \mbox{N = -1, N = 0 } 
    \end{cases}
\end{equation} \par
Having $F(N) = a_1\dots a_p$ where $a_j$ is the bit at position $j$ of the binary representation of $F(N)$, the Fibonacci binary coding is defined as:
\begin{equation}
    F_E(N) = \sum_{i=0}^{j}a_i\cdot F(i)
\end{equation}
Where $a_i \in \{0,1\}$ is the i-th bit of the binary representation of $F(N)$ \cite{walder_ffaasstt_nodate}.

\subsubsection{Delta encoding, Run-length, and Huffman (DRH)}
This technique combines together three well-known compression techniques \cite{mogahed_development_2018}: delta encoding, run-length encoding, and Huffman code. Since these techniques are all lossless, the compression provided by DRH is lossless too if no quantization is applied. \par
Algorithm \ref{ls:DRH_node} describes the compression algorithm, where $Q$ is the quantization level and is asserted to be $\geq 1$: if $Q = 1$ the compression is lossless, and increasing values of $Q$ return higher compression factor and lower reconstruction accuracy.

The decompression algorithm is the inverse of the compression one: once data is received, it is decoded using the Huffman code and reconstructed using the repetition counter. \par 
Since this kind of algorithm is not computationally expensive, the compression phase can be performed also by low resource computational units, such as sensor nodes.

\subsubsection{Sprintz}
Sprintz algorithm \cite{blalock_sprintz_2018} is designed for the IoT scenario, in which energy consumption and speed are important factors. In particular, the goal is to satisfy the following requirements:
\begin{itemize}
	\item Handling of small blocks size
	\item High decompression speed
	\item Lossless data reconstruction
\end{itemize}
The proposed algorithm is a coder that exploits prediction to achieve better results. In particular, it is based on the following components:
\begin{itemize}
	\item \textbf{Forecasting}: used to predict the difference between new samples and the previous ones through delta encoding or FIRE algorithm \cite{blalock_sprintz_2018};
	\item \textbf{Bit packing}: packages are composed of a payload that contains prediction errors and a header that contains information that is used during reconstruction;
	\item \textbf{Run-length encoding}: if a sequence of correct forecasts occurs, the algorithm does not send anything until some error is detected and the length of skipped zero error packages is added as information;
	\item \textbf{Entropy coding}: package headers and payloads are coded using Huffman coding, presented in Subsection~\ref{sub:HC}.
\end{itemize}

\subsubsection{Run-Length Binary Encoding (RLBE)}
\begin{figure}[ht!]
    \centering
    \includegraphics[width=300px]{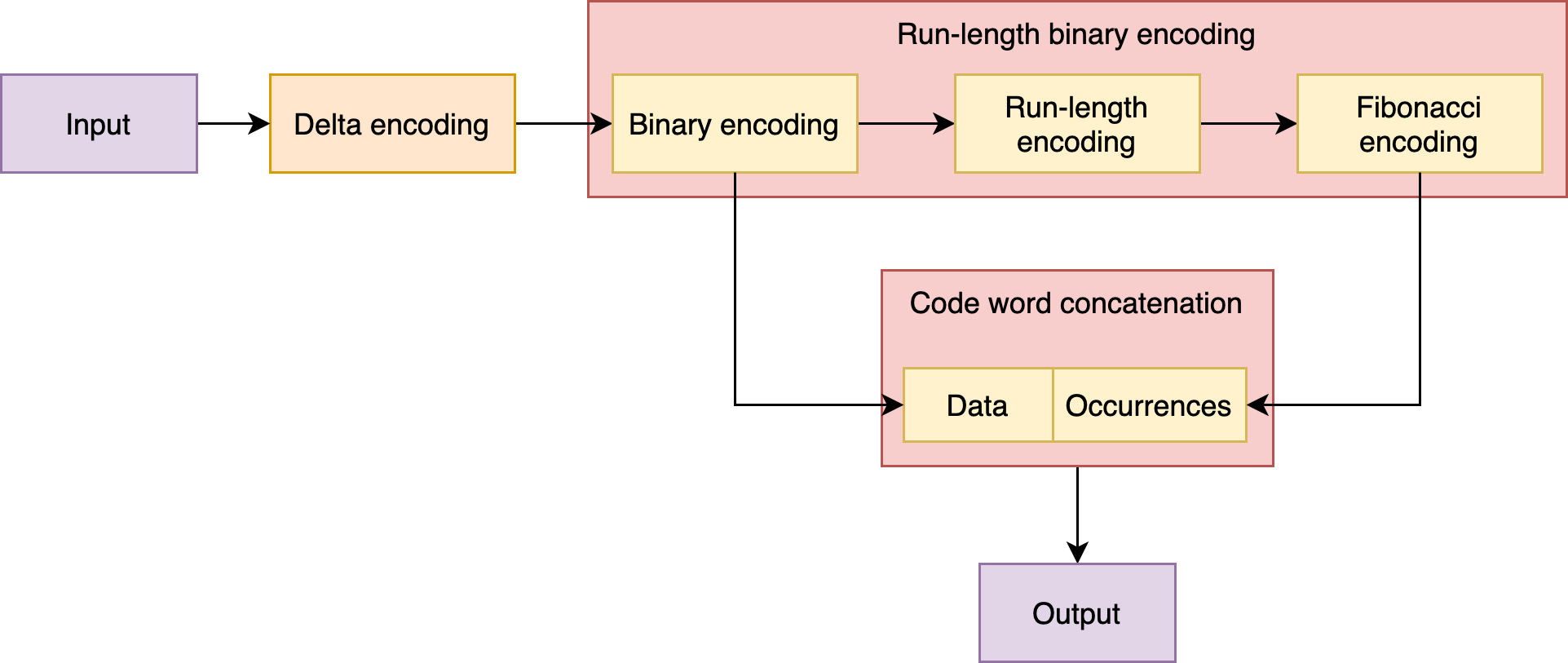}
    \caption{RLBE compression steps}
    \label{fig:RLBE}
\end{figure}

This lossless technique is composed of 5 steps, combining delta encoding, run-length, and Fibonacci coding, as shown in Figure~\ref{fig:RLBE} \cite{spiegel_comparative_2018}.
 \par 
This technique is developed specifically for devices characterized by low memory and computational resources, such as IoT infrastructures.

\subsubsection{RAKE}
\label{subsub:rake}
RAKE algorithm, presented in \cite{campobello_rake_2017}, exploits sparsity to achieve compression. It is a lossless compression algorithm with two phases: preprocessing and compression.

\textbf{Preprocessing} 
In this phase, a dictionary is used to transform original data. For this purpose, many algorithms can be used, such as the Huffman coding presented in Subsection~\ref{sub:HC}, but since the aim is that of obtaining sparsity, the RAKE dictionary uses a code similar to unary coding thus every codeword has at most one bit set to 1. This dictionary doesn't depend on symbol probabilities, so no learning phase is needed. Table~\ref{table:rakeDict} shows a simple RAKE dictionary.
\begin{table}
\centering
\caption{RAKE dictionary}
\label{table:rakeDict}
\begin{tabular}{ccc}
symbol & code          & length    \\ 
-1     & 1             & 1         \\
+1     & 01            & 2         \\
-2     & 001           & 3         \\
+2     & 0001          & 4         \\
...    & ...           & ...       \\
-R     & 0...1         & 2R - 1    \\
+R     & 00...1        & 2R        \\
0      & all zeros     & 2R        \\ 
\end{tabular}%
\end{table}

\begin{figure}[ht!]
	\centering
	\includegraphics[width=218px]{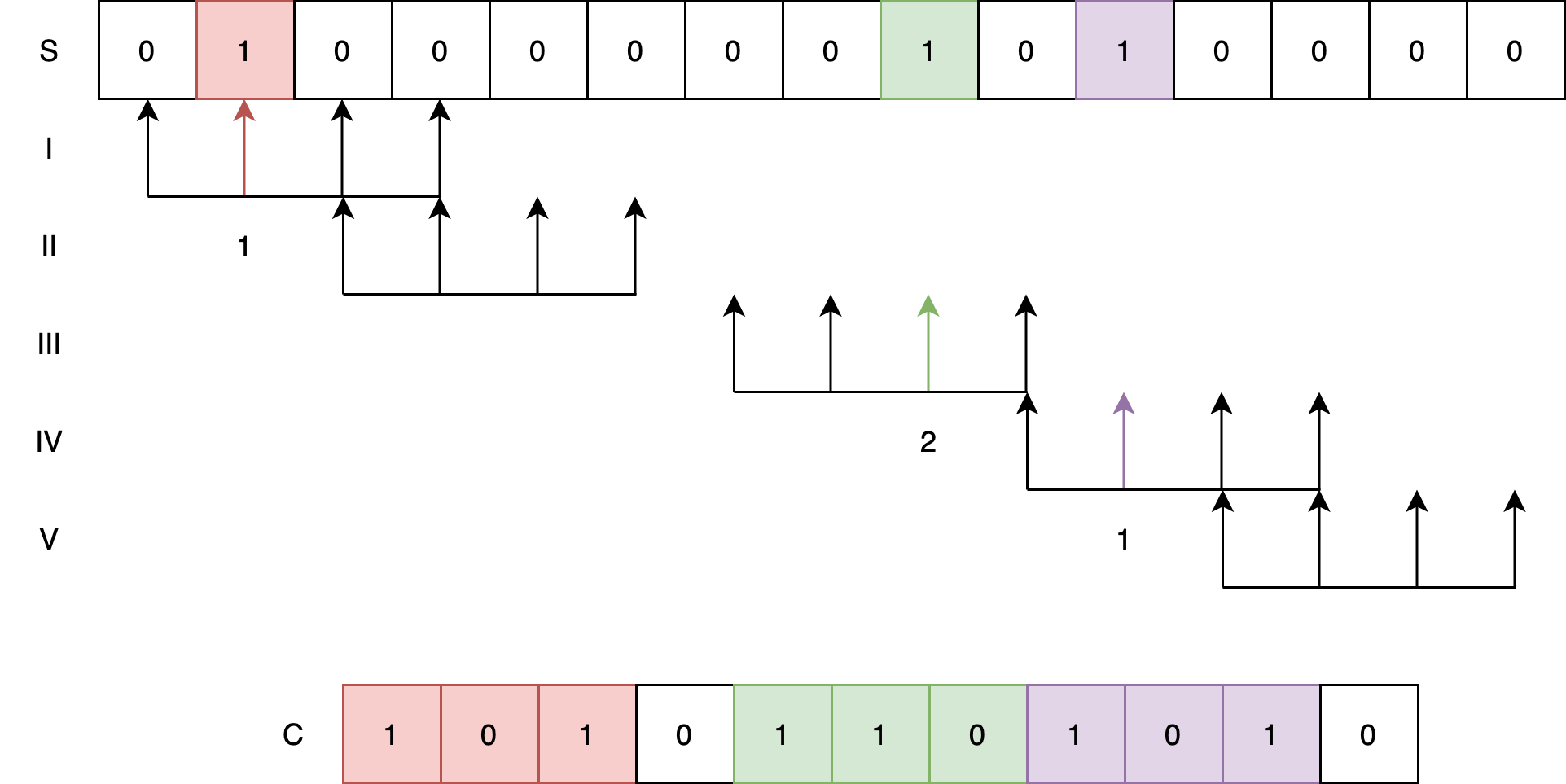}
	\caption{RAKE algorithm execution}
	\label{fig:RAKE}
\end{figure}

\textbf{Compression}
This phase works, as suggested by the algorithm name, as a rake of $n$ teeth. Figure~\ref{fig:RAKE} shows an execution of the compression phase, given a preprocessed input and $n=4$.

The rake starts at position 0, in case there is no bit set to 1 in the rake interval, then 0 is added to the code, otherwise 1 and followed by the binary representation of relative index for the first bit set to 1 in the rake (2 bits in the example in Figure~\ref{fig:RAKE}). After that, the rake is shifted right of $n$ positions for the first case or starting right to the first found bit. In the figure, the rake is initially at position 0, and the first bit set to 1 is in relative position 1 (output: 1 followed by 01), then the rake advances of 2 positions (after the first 1 in the rake); all bits are set to zero (output: 0) thus that rake is moved forward by 4 places; the first bit set to 1 in the rake has relative index 2 (output: 1 followed by 10) thus the rake is advanced by 3 places and the process continues for the two last rake positions (output: 101 and 0 respectively). 

\textbf{Decompression}
The decompression processes the compressed bit stream replacing 0 with $n$ occurrences of 0 and 1 followed by an offset with a number 0 equal to the offset followed by a bit set to 1. The resulting stream is decoded on the fly using the dictionary.  
\subsection{Others}\label{subsec:other}
In this subsection, we present other time series compression algorithms that cannot be grouped in the previously described categories.

\subsubsection{Major Extrema Extractor (MEE)}
This algorithm is introduced in \cite{fink_compression_2011} and exploits time series features (maxima and minima) to achieve compression. For this purpose, strict, left, right, and flat extrema are defined. Considering a time series $TS = [(t_0, x_0), \dots, (t_n,x_n)]$, $x_i$ is a minimum if it follows these rules:
\begin{itemize}
	\item \textbf{strict}: if $x_i < x_{i-1} \land x_i < x_{i+1}$
	\item \textbf{left}: if $x_i < x_{i-1} \land \exists k > i : \forall j \in [i, k], x_j = x_i \lor x_i < x_{k+1}$
	\item \textbf{right}: if $x_i < x_{i+1} \land \exists k < i : \forall j \in [k, i], x_j = x_i \lor x_i < x_{k-1}$
	\item \textbf{flat}: if ($\exists k > i : \forall j \in [i, k], x_j = x_i \lor x_i < x_{k+1}) \land (\exists k < i : \forall j \in [k, i], x_j = x_i \lor x_i < x_{k-1})$
\end{itemize}
For maximum points, they are defined similarly. \par 
\sloppy
After defining minimum and maximum extrema, the authors introduce the concept of importance, based on a distance function \textit{dist} and a compression ratio $\rho$: $x_i$ is an important minimum if:
\begin{eqnarray*}
 \exists i_l < i < i_r : & x_i \text{ is minimum in } \{x_l, \dots, x_r\}\land \\ & \textit{dist}(x_l, x_i) < \rho \land \textit{dist}(x_i, x_r) < r
\end{eqnarray*}
Important maximum points are defined similarly. \par 
Once important extrema are found, they are used as a compressed representation of the segment. This technique is a lossy compression technique, as the Segment Merging one described in the next section. Although the compressed data can be used to obtain original data properties useful for visual data representation (minimum and maximum), it is impossible to reconstruct the original data.

\subsubsection{Segment Merging (SM)}
\label{sub:SM}
\begin{figure}[ht!]
	\includegraphics[width=268px]{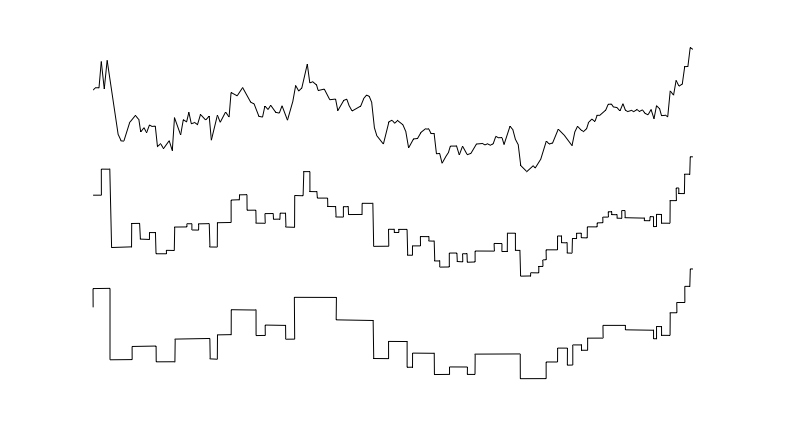}
	\caption{Example of segment merging technique, with increasing value time interval}
	\label{fig:SMExample}
\end{figure}

This technique, presented in \cite{keogh_locally_nodate} and reused in \cite{goldstein_real-time_nodate} and \cite{10.1145/882082.882086}, considers time series with regular timestamps and repeatedly replaces sequences of consecutive elements (segments) with a summary consisting of a single value and a representation error, as shown in Figure~\ref{fig:SMExample} where the error is omitted.
 \par 
After compression, segments are represented by tuples $(t, y, \delta)$ where $t$ is the starting time of the segment, $y$ is the constant value associated with the segment, and $\delta$ is the segment error. The merging operation can be applied either to a set of elements or to a set of segments, to further compress  a previously compressed time series. The case of elements is an instance of the case of segments, and it is immediate to generalize from two to more elements/segments. We limit our description to the merge of two consecutive segments represented by the tuples $(t_i, y_i, \delta_i)$ and $(t_j, y_j, \delta_j)$, where $i < j$, into a new segment represented by the tuple $(t, y, \delta)$ computed as:
\begin{gather*} 
t = t_i  \\ 
y = \frac{\Delta t_i\cdot y_i + \Delta t_j\cdot y_j}{\Delta t_i + \Delta t_j} \\
\delta = \sqrt{\frac{\Delta t_i\cdot (y_i^2 + \delta_i^2) + \Delta t_j\cdot (y_i^2 + \delta_i^2)}{\Delta t_i + \Delta t_j}-y^2} 
\end{gather*}
where $\Delta t_x$ is the duration of the segment $x$, thus $\Delta t_i = t_j - t_i$, $\Delta t_j = t_k - t_j$ and $t_k$ is the timestamp of the segment after the one starting at $t_j$. \par 
The sets of consecutive segments to be merged are chosen to minimize segment error with constraints on a maximal acceptable error and maximal segment duration. \par 
This compression technique is lossy and the result of the compression phase can be considered both as the compressed representation and as the reconstruction of the original time series, without any additional computations executed by a decoder. 

\subsubsection{Continuous Hidden Markov Chain (CHMC)}
The idea behind this algorithm is that the data generation process follows a probabilistic model and can be described with a Markov chain \cite{inamura_keyframe_2003}. This means that a system can be represented with a set of finite states $S$ and a set of arcs $A$ for transition probabilities between states. \par 
Once the hidden Markov chain is found using known techniques, such as the one presented in \cite{young_the_2000}, a lossy reconstruction of the original data can be obtained by following the chain probabilities.
\subsection{Algorithms summary}
\label{subsec:algo_summary}
When we apply the taxonomy described in Section~\ref{sub:compression} and Section~\ref{sec:compressionTechniques} to the above techniques we obtain the classification reported in Table~\ref{table:comparison} that summarizes the properties of the different implementations. To help the reader in visually grasping the membership of the techniques to different parts of the taxonomy, we report the same classification graphically in Figure~\ref{fig:venn} using Venn diagrams. 
\par

\begin{table}
\caption{Compression algorithms classification}
\begin{tabular}{cccccc}
\multicolumn{6}{c}{Dictionary based techniques} \\       
& Non-adaptive & Lossless & Symmetric & min $\rho$ & max $\epsilon$ \\ 
TRISTAN &    -         &      -   & \checkmark  &  \checkmark  & \checkmark\\ 
CORAD  &   -          &      -   & \checkmark  & -   & \checkmark\\
A-LZSS  &-&\checkmark&\checkmark&-&NA\\
D-LZW  &-&\checkmark&\checkmark&\checkmark&NA\\ 
\multicolumn{6}{c}{Function Approximation}  \\ 
& Non adaptive & Lossless & Symmetric & min $\rho$ & max $\epsilon$ \\
PPA     &\checkmark&-&-&\checkmark&\checkmark\\ 
CPT     &\checkmark&-&-&\checkmark&\checkmark\\ 
DWT     &\checkmark&-&-&\checkmark&\checkmark\\ 
DFT     &\checkmark&-&-&\checkmark&\checkmark\\ 
DCT     &\checkmark&-&-&\checkmark&\checkmark\\ 
\multicolumn{6}{c}{Autoencoders}                        \\
& Non adaptive & Lossless & Symmetric & min $\rho$ & max $\epsilon$ \\
RNNA    &-&-&\checkmark&-&\checkmark\\ 
RCA    &-&-&\checkmark&-&\checkmark\\ 
DZip    &-&\checkmark&\checkmark&-&NA\\ 
\multicolumn{6}{c}{Sequential algorithms}               \\
& Non adaptive & Lossless & Symmetric & min $\rho$ & max $\epsilon$ \\ 
DRH     &-&\checkmark&\checkmark&-&NA\\ 
SPRINTZ &\checkmark&\checkmark&\checkmark&-&NA\\
RLBE    &\checkmark&\checkmark&\checkmark&-&NA\\ 
RAKE    &\checkmark&\checkmark&\checkmark&-&NA\\ 
\multicolumn{6}{c}{Others}                    \\
& Non adaptive & Lossless & Symmetric & min $\rho$ & max $\epsilon$ \\
MEE     &\checkmark&-&-&\checkmark&-\\ 
SM      &\checkmark&-&-&\checkmark&-\\ 
CHMC    &-&-&\checkmark&-&-\\
\end{tabular}%
\label{table:comparison}
\end{table}
Where $\checkmark$ indicates if the related property is true, - if it is not, and NA if it is not applicable. Min $\rho$ and max $\epsilon$ indicate respectively the possibility to set a minimum compression ratio or a maximum reconstruction error.
\section{Experimental results} \label{sec:experimentalResults} In this section, we discuss the different performances of the techniques in Section~\ref{sec:compressionTechniques}, as reported by their authors. To ensure a homogeneous presentation of the different techniques and for copyright reasons, we do not include figures from the works we are describing. Instead, we redraw the figures using the same graphical style for all of them and maintain a faithful reproduction with respect to the represented values. \par 

The experiments presented in those studies where based on the the following datasets:
\begin{itemize}
	\item ICDM challenge \cite{wojnarski_IEEE_2010} \--- Traffic flows;
	\item RTE France \footnote{https://data.rte-france.com} \--- Electrical power consumptions;
	\item ACSF1\footnote{http://www.timeseriesclassification.com} \--- Power consumptions;
    \item BAFU\footnote{https://www.bafu.admin.ch}, \---  Hydrogeological data;
	\item GSATM\footnote{\label{UCI}https://archive.ics.uci.edu} \--- Dynamic mixtures of carbon monoxide (CO) and humid synthetic air in a gas chamber;
	\item PigAirwayPressure\footnote{\label{UCR}https://www.cs.ucr.edu/}\--- vital signs;
	\item SonyAIBORobotSurface2\textsuperscript{\ref{UCR}} \--- X-axis of robot movements;
	\item SMN \cite{smith_the_2008} \---  Seismic measures in Nevada;
	\item HAS \cite{shoaib_fusion_2014} \---  Human activity from smartphone sensors;
	\item PAMAP \cite{reiss_towards_2011} \--- Wearable devices motion and heart frequency measurements;
	\item MSRC-12\footnote{https://www.microsoft.com}
	\--- Microsoft Kinect gestures;
	\item UCI gas dataset\textsuperscript{\ref{UCI}} \--- Measuring gas concentration during chemical experiments;
	\item AMPDs\footnote{http://ampds.org} \--- Measuring maximum consumption of water, electricity, and natural gas in houses.
\end{itemize}

\subsection{Algorithms}
To foster the comparison of the different algorithms, we report synoptically the accuracy and compression ratio obtained in the experiments described by their authors.

\subsubsection{TRISTAN}

\begin{figure}[ht!]
	\centering
	\includegraphics[width=218px]{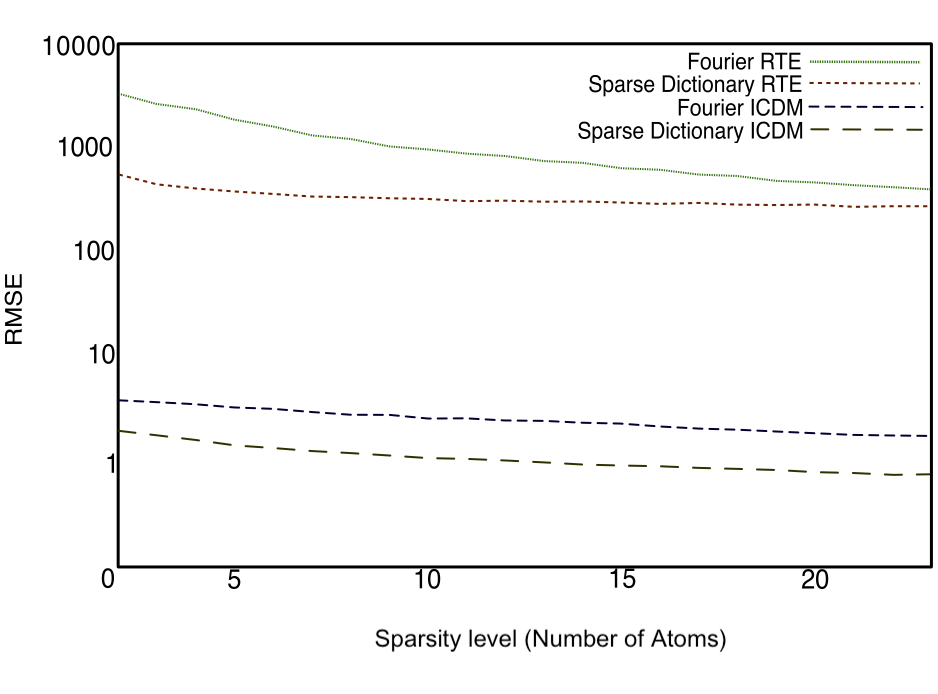}
	\caption{RMSE results depending on sparsity for TRISTAN\cite{marascu_tristan_2014}}
	\label{fig:TristanRMSE}
\end{figure}

\textbf{Accuracy}
The creation of the dictionary depends on the dictionary sparsity parameter that is directly related to its size and the accuracy can be influenced by this value, as shown in Equation~\ref{eq:TRISTAN-dictionary}. Experimental results are shown in Figure~\ref{fig:TristanRMSE}, where sparsity is related to RMSE.

\noindent \textbf{Compression ratio}
The sparsity parameter also affects the compression ratio. To assess this dependency, the authors execute experiments on two different datasets both containing one measurement per minute with daily segments: RTE and ICDM, using dictionaries consisting respectively of 16 and 131 atoms. 

The resulting compression ratios are $\rho = 0.5$ for RTE and $\rho = 0.05$ for ICDM, thus higher sparsity values (larger dictionaries) allow for better compression ratios.

\subsubsection{CORAD}
\textbf{Accuracy}
The authors of CORAD measure accuracy using MSE and the best-reported results for their algorithms are 0.03 for the BAFU dataset, 0.11 for the GSATM dataset, and 0.04 for the ACSF1 dataset. As for TRISTAN, accuracy depends on time series segments redundancy and correlation.

\noindent \textbf{Compression ratio}
The best reported results for CORAD compression ratio are $\rho=0.04$ for the ACSF1 dataset, $\rho = 0.07$ for the BAFU dataset and $\rho = 0.06$ for the GSATM dataset. \par
\begin{figure}
  \centering
  \subfloat[Varying error threshold]{\includegraphics[width=200px]{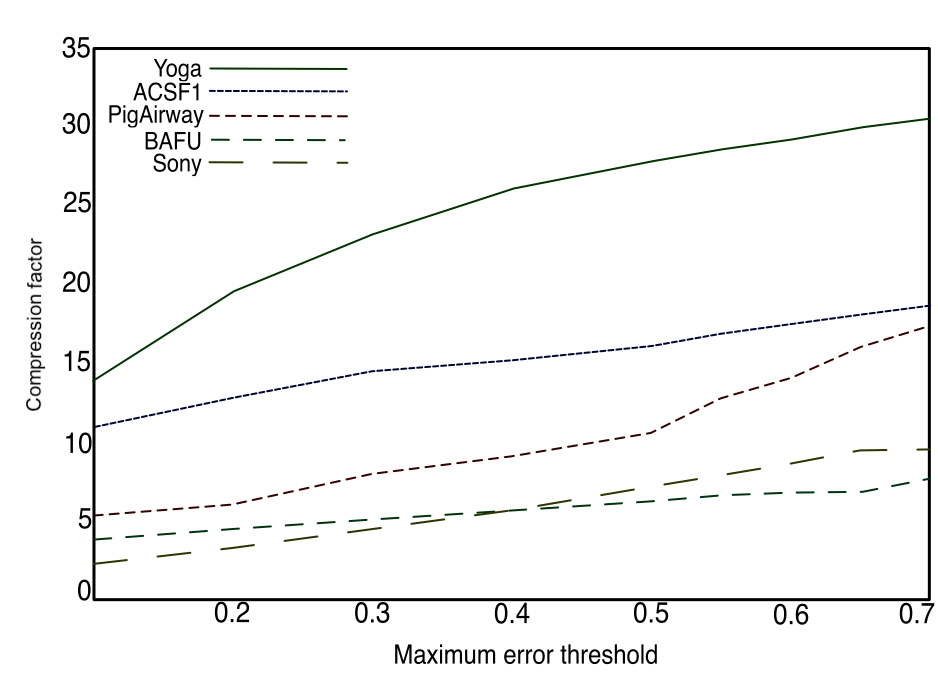}} 
  \par\bigskip
  \subfloat[Varying dictionary atoms]{\includegraphics[width=200px]{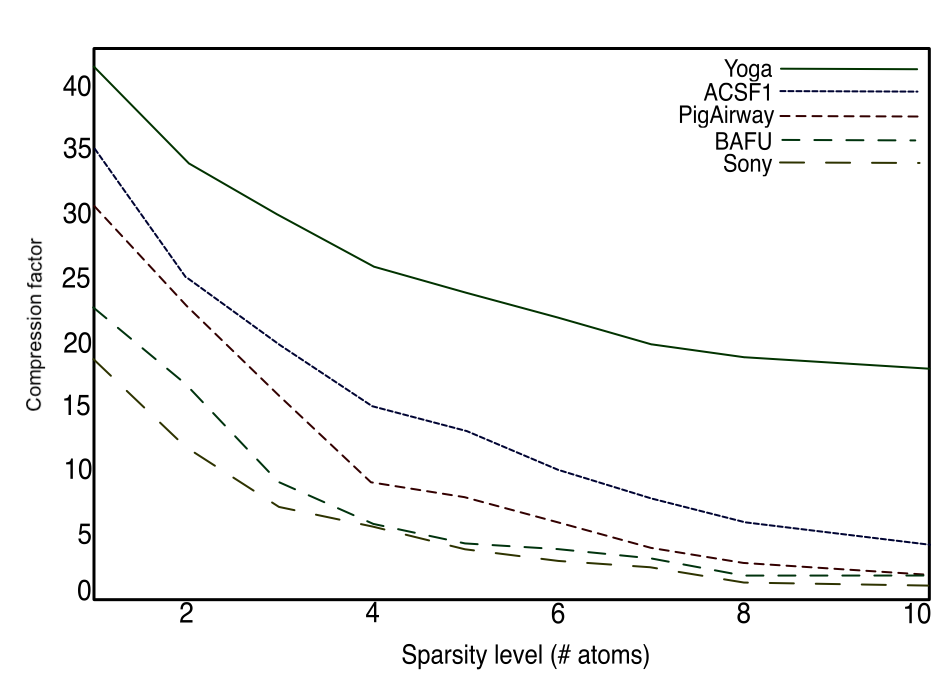}}
  \par\bigskip
   \subfloat[Varying segment length]{\includegraphics[width=200px]{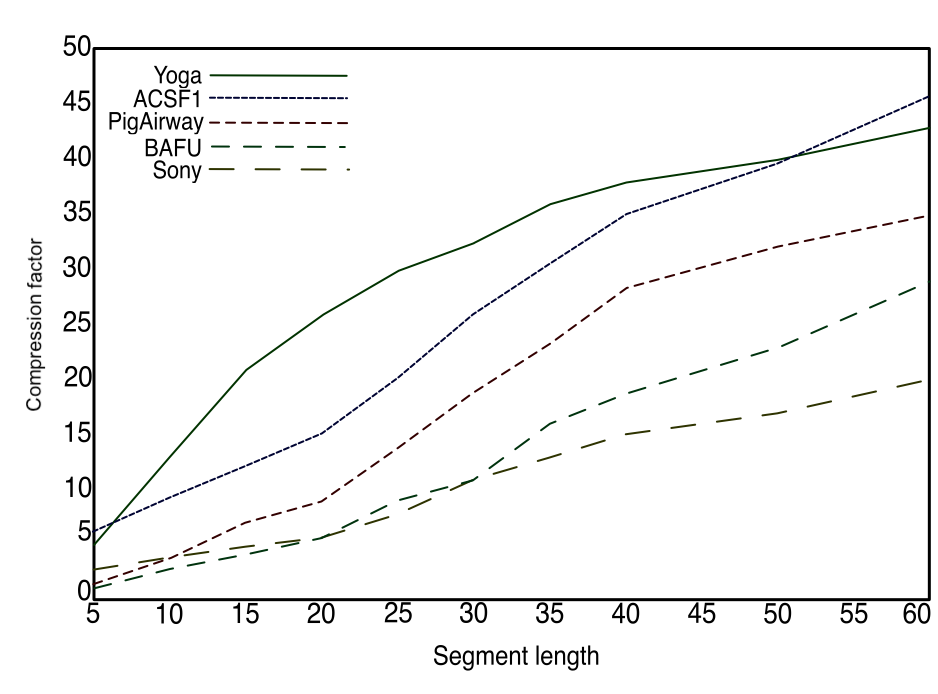}}
  \caption{Compression factor ($\frac{1}{\rho}$) with varying parameters for CORAD \cite{khelifati_corad_nodate}}
  \label{fig:compressionratioconrad}
\end{figure}

The compression ratio is affected by the parameters used during training and encoding: error threshold, sparsity, and segment length. Their effects are depicted in the three plots in Figure~\ref{fig:compressionratioconrad}, that represent the compression factor ($\frac{1}{\rho}$) obtained for different values of the parameters.
 \par

\subsubsection{A-LZSS}

The compression ratio obtained by the A-LZSS algorithm depends on two parameters: $L_n$ and $D_n$, which are respectively the lookahead distance, and the dictionary atoms length. From the experiments conducted by the authors in \cite{pope_accelerometer_2018}, the best compression ratio is obtained setting $L_n=5$ and $D_n=10$. Increasing $L_n$ and decreasing $D_n$ penalize this quality index, as values of $L_n < 4$.

\subsubsection{D-LZW}

The authors didn't provide a detailed study on the compression performances, limiting the experimental results only to EGC and PPG signals. Nevertheless, they did a comparison with Bzip2 and Gzip, giving the results shown in Table \ref{tab:DLZWResults}.

\begin{table}[ht!]
    \centering
    \caption{D-LZW compression ratio comparison}
\begin{tabular}{cccc}
                       & D-LZW & Bzip2 & Gzip  \\
Compression ratio (\%) & 76.47 & 60.52 & 53.63
\end{tabular}
    
    \label{tab:DLZWResults}
\end{table}
 
\subsubsection{Piecewise Polynomial Approximation (PPA)}
Accuracy and compression ratio of PPA are affected by the maximum polynomial degree and the maximum allowed deviation, the parameters of the method as described in Subsection~\ref{subsec:PPA}.

\noindent\textbf{Accuracy}

Errors are bounded by the maximum deviation parameter and the maximum polynomial degree can affect accuracy: deviation is always under the threshold and using higher degree polynomials yield more precise reconstructions. \par
\begin{figure}[ht!]
  \centering
    \includegraphics[width=200px]{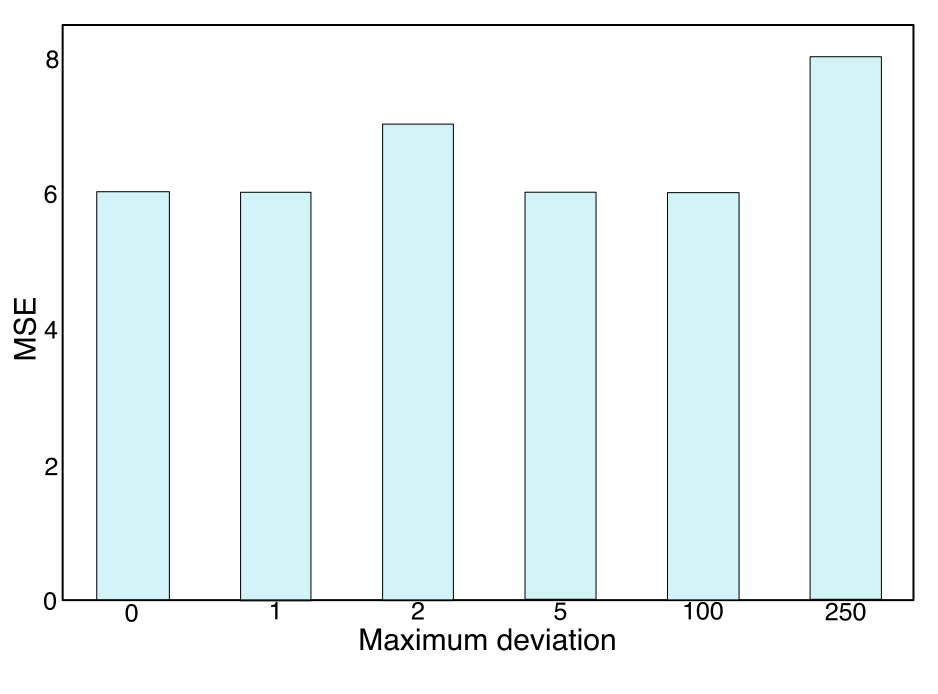}
    \caption{Relation between MSE and maximum deviation for PPA\cite{hsu_time_2017}}
  \label{fig:PPAMSA}
\end{figure}

 
Figure~\ref{fig:PPAMSA} shows the relation between maximum deviation and MSE.

\noindent \textbf{Compression ratio}

\begin{figure}[ht!]
  \centering
  \subfloat[Varying maximum polynomial degree]{\includegraphics[width=200px]{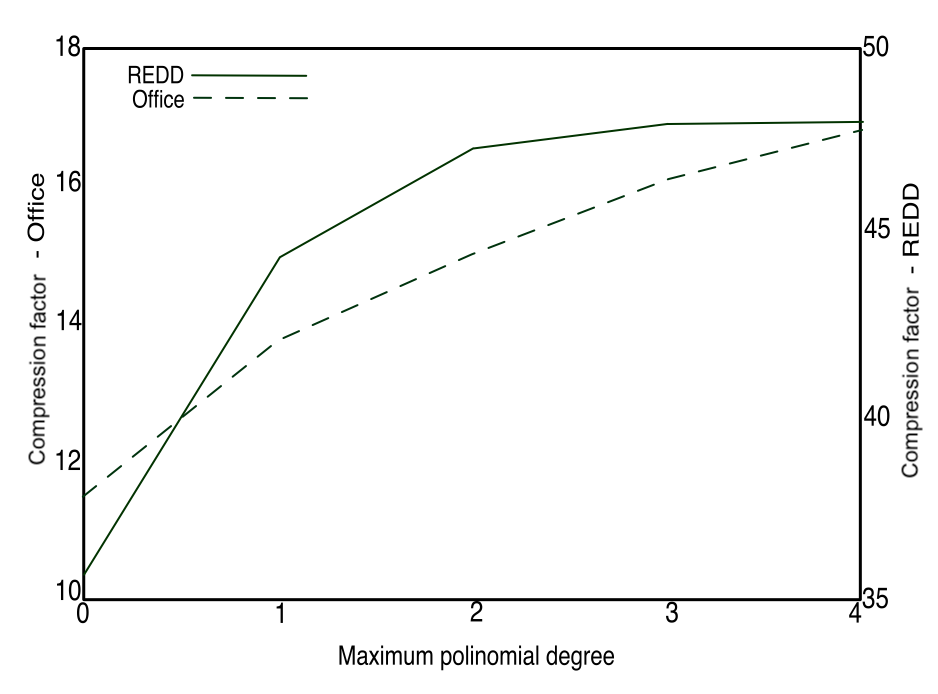}} 
  \par\bigskip
  \subfloat[Varying maximum deviation]{\includegraphics[width=200px]{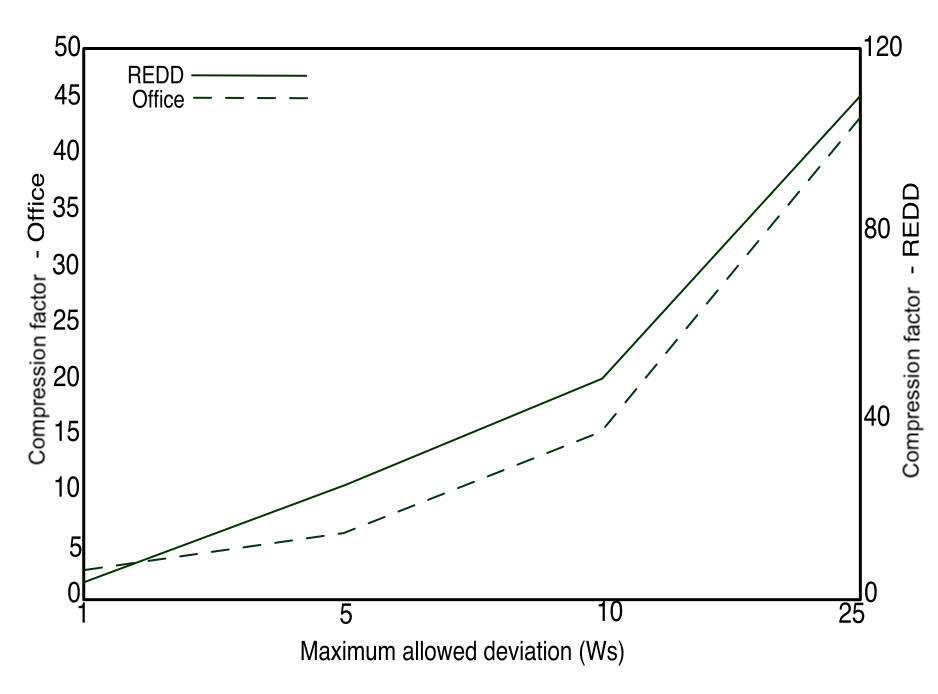}} 
  \par\bigskip
  \caption{PPA \cite{eichinger_time-series_2015} compression factor ($\frac{1}{\rho}$) for varying parameters}
  \label{fig:compressionratioppa}
\end{figure}

Both the maximum deviation and polynomial degree parameters affect the compression ratio, as we can observe in Figure~\ref{fig:compressionratioppa} where the compression factor ($\frac{1}{\rho}$) is reported for different values of parameters. 
The best compression ratios, $\rho = 0.06$ for REDD and $\rho = 0.02$ for Office, are achieved for the highest degree polynomials. As expected, also for increasing maximum deviation the compression factor is monotonically increasing.

\subsubsection{Chebyshev Polynomial Transform (CPT)}
Despite the absence of experimental results in the original paper, some considerations can be done. The algorithm has three parameters:
\begin{itemize}
    \item Block size;
    \item Maximum deviation;
    \item Number of quantization bits.
\end{itemize}
Similarly to PPA, in CPT it is possible to obtain higher compression ratios at the cost of higher MSE by increasing the maximum deviation parameter. The same holds for the block size: larger blocks correspond to better compression ratios and higher decompression errors \cite{hawkins_algorithm_2012}. Similarly, we can suppose that using fewer quantization bits would entail a less accurate decompression and a better compression ratio.

\subsubsection{Discrete Wavelet Transform (DWT)}
As for CPT, experimental results are not provided, but also DWT performances follow the same rules as for PPA since the compression ratio and threshold errors are strictly correlated. The threshold error can be fixed a priori to have a higher compression ratio or a higher accuracy.

\subsubsection{Discrete Fourier Transform (DFT)}

The authors tested the performance of the DFT algorithm using three types of signals: Block, Heavy Sine, and Mishmash. The compression ratio depends on the percentage of coefficients that are removed. The optimum threshold is derived by iterative tests, and it depends on the signal type. As stated by the authors, the results obtained by DWT are comparable to the ones obtained by DWT: the choice of the algorithm highly depends on the type of signal that has to be compressed \cite{karim_wavelet_2011}.

\subsubsection{Discrete Cosine Transform (DCT)}

The accuracy obtained by this algorithm depends on the chosen compression ratio. The authors used the ARIANE 5 (temperature sensor, and vibration sensor) and AISat temperature sensor for their experiments. The results show that the performances of this algorithm are not homogeneous, as they highly depend on the input dataset. Even if the input data seems to be similar, belonging to the same domain: the performances obtained on the ARIANE 5 and AISat temperature sensors are complitely different, as shown in Table \ref{tab:DCTRES} \cite{mess_compression_nodate}.

\begin{table}[ht!]
\caption{DCT experimental results}
\begin{tabular}{lcc}
\multicolumn{1}{c}{}                                       & Compression ratio (\%) & MSE    \\
\multicolumn{1}{c}{{A5 temperature sensor}} & 49.60                  & 0.0009 \\
\multicolumn{1}{c}{}                                       & 82.60                  & 0.03   \\
{A5 vibration sensor}                       & 3.10                   & 0.04   \\
                                                           & 7.10                   & 0.10   \\
                                                           & 22.90                  & 0.30   \\
{AISat temperature sensor}                  & 2.30                   & 6.50   \\
                                                           & 5.20                   & 92.80  \\
                                                           & 8.80                   & 317.92
\end{tabular}
\label{tab:DCTRES}
\end{table}

\subsubsection{Recurrent Neural Network Autoencoder (RNNA)}

\textbf{Accuracy}
One of the parameters of the RNNA based methods is the threshold for the maximally allowed deviation, whose value directly affects accuracy.
For the experiments on univariate time series presented in \cite{hsu_time_2017} the authors report, considering the optimal value for deviation parameter, an RMSE value in the range $[0.02, 0.07]$ for different datasets. In the same experiments, RMSE is in the range $[0.02, 0.05]$ for multivariate time series datasets. Thus, there is no significant difference between univariate and multivariate time series.

\noindent \textbf{Compression ratio}
Similarly, the maximally allowed deviation affects the compression ratio. According to the results reported by the authors, in this case, there is not a significant difference between univariate and multivariate time series: $\rho$ is in the range $[0.01,0.08]$ for the univariate time series dataset and in the range, $[0.31,0.05]$  for the multivariate one. 

\subsubsection{Recurrent Convolutional Autoencoder (RCA)}

The experimental results reveal a slight improvement respect to the RNNA model over the traffic flow dataset\cite{zhao_time_2019}: without the convolutional layer, the RMSE value is 10.37, while with the convolutional layer, the RMSE decreases to 8.17. 

Since the original paper is focused on the prediction of the next values only, the compression ratio is missing.

\subsubsection{DZip}

The authors tested the model with real and synthetic datasets and compared the results with the state-of-the-art compression algorithms, including RNNA. The results revealed that it outperforms RNNA autoencoder. The compression ratio depends on the vocabulary size, that is negatively affected by larger sizes. Another factor that influences this parameter is the size of the original data: larger is the time series to be compressed, better is the compression ratio \cite{goyal_dzip_2021}.

\subsubsection{Delta encoding, Run-Length, and Huffman (DRH)}
DRH, as most of the following algorithms, is a lossless algorithm, thus the only relevant performance metric is the compression ratio. \par 
\noindent \textbf{Compression ratio} This parameter highly depends on the dataset: run-length algorithm combined with delta encoding achieves good results if the time series is composed of long segments that are always increasing or decreasing of the same value or constant.
Experimental results are obtained on temperature, pressure, and light measures datasets. For this type of data, DRH appears to be appropriate, since the values are not highly variable and have a relatively long interval with constant values. The resulting compression factors are reported in Table~\ref{tab:DRHCR}  \cite{mogahed_development_2018}.
\begin{table}[ht!]
\caption{DRH Compression factors}
\centering
\begin{tabular}{cc}
                    & $^1/_\rho$ \\ 
Indoor temperature  & 92.2              \\ 
Outdoor temperature & 91.17             \\ 
Pressure            & 82.55             \\ 
Light               & 83.53             \\ 
\end{tabular}
\label{tab:DRHCR}
\end{table}

\subsubsection{Sprintz}

\begin{figure}[ht!]
    \centering
    \includegraphics[width=200px]{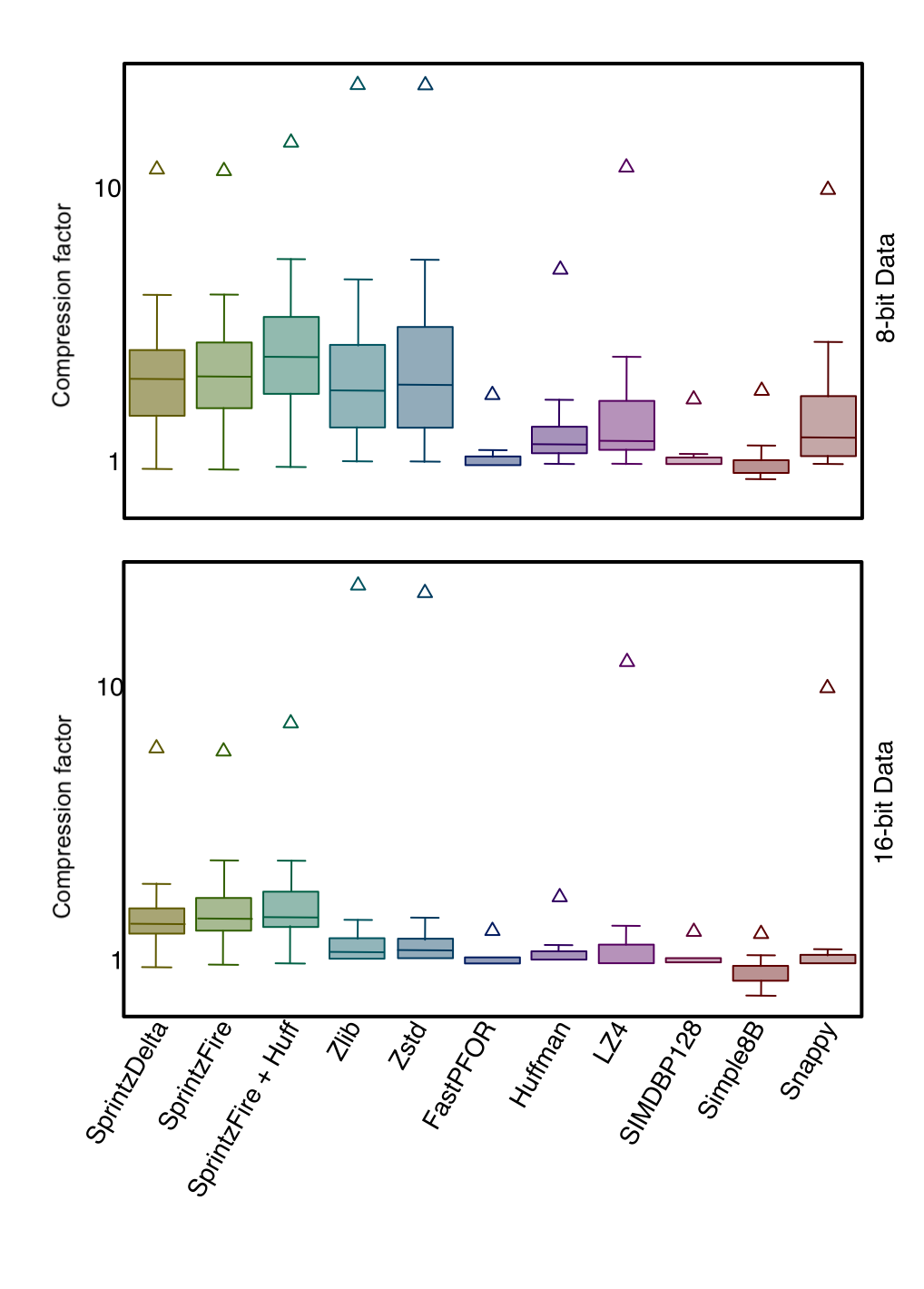}
    \caption{Sprintz compression factors versus other lossless compression algorithms \cite{blalock_sprintz_2018}}
    \label{fig:SprintzCR}
\end{figure}

\noindent \textbf{Compression ratio} This algorithm is tested over the datasets of the UCR time series archive, which contains data coming from different domains. In the boxplots in Figure~\ref{fig:SprintzCR}, compression factors are shown for 16 and 8-bit data and compared with other lossless compression algorithms.
From this graph, it is possible to see that all algorithms achieve better (higher) compression factors on the 8-bit dataset. This could be due to the fact that 16-bit data are more likely to have higher differences between consecutive values. \par 
Another interesting result is that the FIRE forecasting technique improves this compression, especially if combined with Huffman coding when compared with Delta coding \cite{blalock_sprintz_2018}.

\subsubsection{Run-Length Binary Encoding (RLBE)}

\begin{figure}[ht!]
    \centering
    \includegraphics[width=200px]{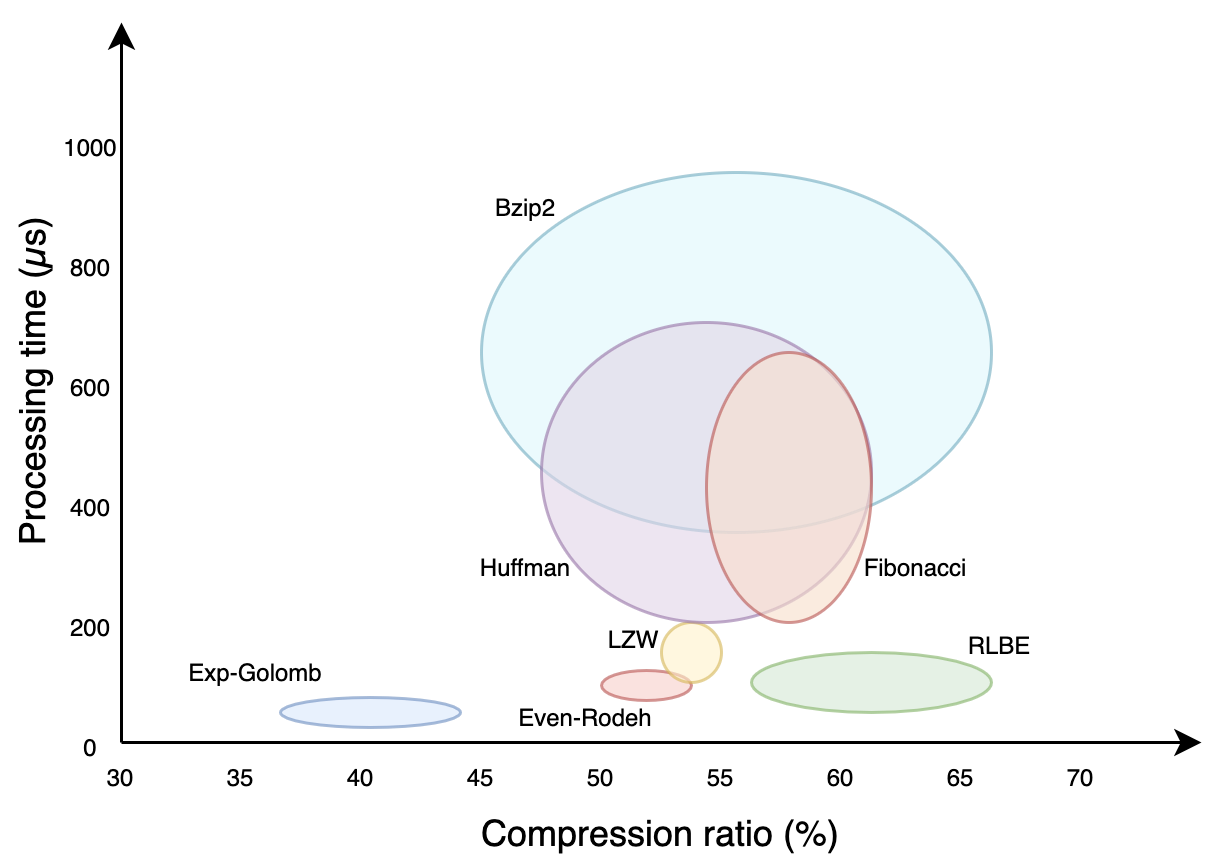}
    \caption{RLBE compression ratio versus other lossless compression algorithms \cite{spiegel_comparative_2018}}
    \label{fig:RLBECR}
\end{figure}

\noindent \textbf{Compression ratio} 
The author of this algorithm in~\cite{spiegel_comparative_2018} report compression ratio and processing time of RLBE with respect to those of other algorithms as shown in Figure~\ref{fig:RLBECR} where ellipses represent the inter-quartile range of compression ratios and processing times. RLBE achieves good compression ratios with a 10\% variability and with low processing time.
\par 

\subsubsection{RAKE}

\noindent \textbf{Compression ratio}
The authors of RAKE in~\cite{campobello_rake_2017} run experiments on sparse time series with different sparsity ratios $p$ and different algorithms. The corresponding compression factors are reported in  
Table~\ref{tab:RAKECR}. 
\begin{table}[ht!]
\caption{Compression factor results for RAKE versus other lossless compression algorithms}

\begin{tabular}{ccccccccc}
          & \multicolumn{8}{c}{p}                                \\ 
Algorithm & 0.002 & 0.005 & 0.01 & 0.05 & 0.1 & 0.15 & 0.2 & 0.25 \\ 
OPT       & 48.0  & 22.0  & 12.4 & 3.5  & 2.1 & 1.6  & 1.4 & 1.2  \\ 
RAKE      & 47.4  & 21.5  & 12.0 & 3.5  & 2.1 & 1.6  & 1.4 & 1.2  \\ 
RAR       & 22.1  & 12.1  & 7.4  & 2.6  & 1.7 & 1.4  & 1.2 & 1.2  \\ 
GZIP      & 21.4  & 11.8  & 7.4  & 2.6  & 1.8 & 1.4  & 1.3 & 1.2  \\ 
BZIP2     & 26.0  & 14.2  & 8.7  & 2.6  & 1.7 & 1.3  & 1.2 & 1.1  \\ 
PE        & 29.5  & 11.9  & 5.9  & 1.2  & 0.6 & 0.4  & 0.3 & 0.2  \\ 
RLE       & 35.7  & 15.7  & 8.4  & 2.1  & 1.2 & 0.9  & 0.8 & 0.6  \\ 
\end{tabular}%
\label{tab:RAKECR}
\end{table} 
We can notice that the compression factor is highly dependent on sparsity and lower sparsity values correspond to higher compression factors (better compression). Thus, the RAKE algorithm is more suitable for sparse time series.

\subsubsection{Major Extrema Extractor (MEE)}
The compression ratio is one of the parameters of the MEE algorithm, thus the only relevant performance metric is the accuracy. \par 
 
\textbf{Accuracy}
Even though MEE is a lossy compression algorithm, the authors have not provided any information about accuracy. Since the accuracy depends on the compression ratio, the authors recommend choosing a value that is not smaller than the percentage of non-extremal points, to obtain an accurate reconstruction of the compressed data \cite{fink_compression_2011}. 

\subsubsection{Segment Merging (SM)}
\begin{figure}[ht!]
    \centering
    \includegraphics[width=200px]{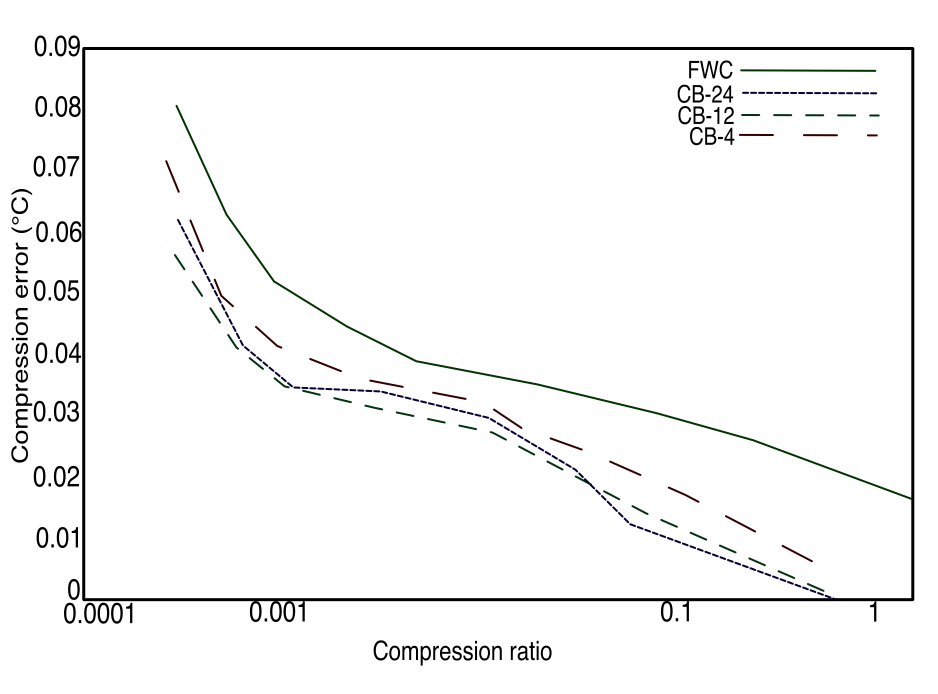}
    \caption{Correlation between compression ratio and error on a temperature dataset \cite{goldstein_real-time_nodate}}
    \label{fig:SMPlot}
\end{figure}

Since this algorithm is used for visualization purposes, a first evaluation can be a qualitative observation. In Figure~\ref{fig:SMExample} we can see how a light sensor time series is compressed considering different time series segment lengths. 
For this algorithm, compression ratio and error are strictly correlated, as shown in Figure~\ref{fig:SMPlot} where FWC and CB-$m$ correspond to different versions of the algorithm and different parameters~\cite{goldstein_real-time_nodate}. 
We can observe that compression ratio values close to 1 correspond to error values close to zero. 

The dependence of compression ratio and error changes for different datasets. 
Knowing the function that correlates the compression ratio and error for a given dataset, it is possible to set the compression ratio that corresponds to the desired maximum compression error.

\subsubsection{Continuous Hidden Markov Chain (CHMC)}
For this algorithm, the authors haven't provided any quantitative evaluation for compression performances: their assessment is qualitative and mainly focused on the compression and reconstruction of humanoid robot motions after a learning phase \cite{inamura_keyframe_2003}.

\subsection{Intergroup comparison}

After passing through the single techniques, we propose an intergroup comparison to address the general characteristics of each group.

\subsubsection{Dictionary-based}

The dictionary-based approaches include both lossy and lossless compression. The compression ratio depends on the atoms size in the dictionary and on the dataset repetitiveness. For large datasets, the storage needed for the dictionary can be considered irrelevant, while for smaller datasets it can be inefficient. The training phase can be inefficient for databases with low repetitiveness.

\subsubsection{Functional approximation}

Functional approximation includes only lossy techniques. The compression ratio can be set in advance, as the maximum reconstruction error. They work better with regular data with low fluctuations and peaks.

\subsubsection{Autoencoders}

Autoencoders includes lossy and lossless techniques. The error threshold can be set in advance, while the compression ratio depends on the quality of the training phase and in the regularity of the dataset. If the autoencoder receives data with different seasonability and distribution respect to the ones found in the training set, its prediction capacity becomes low. A weak point is the computational cost of the training set: recurrent neural networks are more expensive to be trained respect to classic neural network autoencoders and need a large amount of data in the training set.

\subsubsection{Sequential algorithms}

The selected techniques in this group includes lossless techniques only. The compression ratio depends on the data sparsity and regularity, especially for those that use delta and run-length encoding. They don't need a training phase and work well with low-computational resources.

\subsubsection{Others}

This group includes lossy techniques, and their peculiarities depend on the single technique.

\section{Conclusions} \label{sec:Conclusions}
The amount of time series data produced in several contexts requires specialized time series compression algorithms to store them efficiently. In this paper, we provide an analysis of the most relevant ones, we propose a taxonomy to classify them, and report synoptically the accuracy and compression ratio published by their authors for different datasets. Even if this is a meta-analysis, it is informative for the reader about the suitability of a technique for specific contexts, for example, when a fixed compression ratio or accuracy is required, or for a dataset having particular characteristics. 

The selection of the most appropriate compression technique highly depends on the time series characteristics, the constraints required for the compression, or purpose.

Starting from the first case, time series can be characterized by their regularity (that means, having or not fixed time intervals between two consecutive measurements), sparsity, fluctuations, and periodicity. Considering \textbf{regularity}, only techniques based on function approximation can be applied directly to the time series without the need of transformations. This is because that approach approximates a function passing through the given points, having in the x-axis time, so the time-distance between consecutive point can be arbitrary. All the other techniques assumes the time-distance between two points to be fixed. To apply them, some transformations on the original time series are needed. Some techniques work better having in input time series with high \textbf{sparsity}, as the ones proposed in Section \ref{sec:sequentals}. Another characteristic to consider is \textbf{fluctuations}. The lossy compression techniques presented in this survey cannot achieve good accuracy results having in input time series with high non-regular fluctuations. This is caused by the fact that functional approximation and autoencoder techniques tend to flatten peaks, while the ones based on dictionaries would need overly large dictionaries. In the other hand, functional approximation and lossy autoencoders can be taken in consideration if it is sufficient to reconstruct the general trend of the original time series. Lastly, dictionary-based techniques work well with time series with high \textbf{periodicity}. In this scenario, a relatively small dictionary could be enough to represent the whole time series by combining atoms. Lossy autoencoders also improve their reconstruction accuracy, having in input periodical time series.

The constraints required for compression include compression ratio, reconstruction accuracy, speed, and availability of a training dataset. As shown in Table \ref{table:comparison}, some techniques allow to specify a minimum \textbf{compression ratio} and a maximum \textbf{reconstruction accuracy} error. This could be necessary in some contexts in which space or accuracy are given as constraints. Moreover, compression \textbf{speed} can be relevant, in particular when hardware with low computation capacity are involved. For this context, sequential algorithms can compress at high speed, without needing powerful devices. Depending on the presence or not of a \textbf{training set}, all the techniques that need a training phase must be discarded.

The different purposes for time series compression include analysis, classification, and visualization. For time series \textbf{analysis}, functional approximation and lossy autoencoders can help to reduce noise and outlayers, for those time series that have many fluctuations. Autoencoders can be used also to more deep analysis, as anomaly detections, starting from the compressed representation, as shown in \cite{provotar_unsupervised_2019}. The CHMC technique can also be used for this aim, since time series can be represented with a probability graph. Autoencoders can be used for \textbf{classification} too, since the autoencoder can be used to reduce the dimensionality of the input. Lastly, for \textbf{visualization} purposes, MEE and SM can be used to having different definitions of long time series. 

\appendix

\section{Algorithms}

\begin{lstlisting}[caption=Dictionary based - Training, label=code:DB-training, language=Python]
createDictionary(Stream S, Threshold th, int segmentLength) {
    Dictionary d;
    Segment s;
    while (S.isNotEmpty()) {
        s.append(S.read());
        if (s.length == segmentLength) {
            if (d.find(s, th)) {
                d.merge(s);
            } else {
                d.add(s);
            }
            s.empty();
        }
    }
    return d;
}
\end{lstlisting}

\begin{lstlisting}[caption=Dictionary based - Compression, label=code:DB-compression]
compress(Stream S, Dictionary d, Threshold th, int segmentLength) {
    Segment s;
    while (S.isNotEmpty()) {
        s.append(S.read());
        if (s.length == segmentLength) {
            if (d.find(s, th)) {
                send(d.getIndex(s));
            } else {
                send(s);
            }
            s.empty();
        }
    }
}
\end{lstlisting} 

\begin{lstlisting}[caption=A-LZSS algorithm, label=ls:ALZSS]
compress(Stream S, int minM, Dictionary d, int Ln, int Dn) {
    foreach(s in S) {
        I, L = d.longestMatch(s, Ln, Dn);
        if (L < minM) {
            send(getHuffCode(s));
        } else {
            send((I, L));
            s.skip(L);
        }
    }
}
\end{lstlisting}

\begin{lstlisting}[caption=PPA algorithm, label=ls:PPA, mathescape=true]
compress(int $\rho$, Stream S, float $\epsilon$) {
    Segment s = S.read();
    while (S.isNotEmpty()) {
        Polynomial p;
        int l = 0;
        foreach (k in [0 : $\rho$]) {
            float currErr;
            Polynomial currP;
            bool continueSearch = true;
            int i = 0;
            int currL;
            while (continueSearch && i < len(s)) {
                currErr, currP = approx(k, s.getPrefix(i));
                if (currErr < $\epsilon$) {
                    i += 1;
                    currL = i;
                } else {
                    continueSearch = false;
                }
            }
            while (continueSearch && S.isNotEmpty()) {
                s.append(S.read());
                currErr, currP = approx(k, s);
                if (currErr < $\epsilon$) {
                    currL = len(s);
                } else {
                    currL = len(s) $-$ 1;
                    continueSearch = false;
                }
            }
            if (currErr < $\epsilon$ && currL > l) {
                p = currP;
                l = currL;
            }
        }
        if (len(s) > 0) {
            send(p, l);
            s.removePrefix(l);
        } else {
            throw "Error: exceeded error threshold value";
        }
    }
}
\end{lstlisting}

Where:

\begin{itemize}
    \item $S$ is the input time series;
    \item $\rho$ is the maximum polynomial degree;
    \item $\epsilon$ is the error threshold.
\end{itemize}

\begin{lstlisting}[caption=RNN compression algorithm, label=code:RRACA, mathescape=true]
compress(Stream S, float $\epsilon$, RAE a) {
    Segment s = Null;
    while (S.isNotEmpty()) {
        Segment aux = s;
        Element e = S.read();
        aux.append(e);
        if (getError(aux, a.decode(a.encode(aux)) < $\epsilon$) {
            s = aux
        } else {
            send(a.encode(s));
            s.empty();
            s.append(e);
        }
    }
}
\end{lstlisting}

\begin{lstlisting}[caption=Huffman Code compression, label=code:HMCode, mathescape=true, float]
createDictionary(StreamPrefix S) {
    Tree T = new Tree();
    PriorityList L = createPriorityList(S);
    foreach (i in [0 : L.length $-$ 1]) {
        Node n = new Node();
        Element el1 = L.extractMin();
        Element el2 = L.extractMin();
        n.frequency = el1.frequency + el2.frequency;
        T.addTree(n, (el1, el2));
        L.add(n);
    }
    return L.toDictionary();
}

compress(Stream S, int prefixLen) {
    StreamPrefix s = S.prefix(prefixLen);
    Dictionary D = createDictionary(s);
    CompressedRepresentation R = [];
    while (S.isNotEmpty()) {
        Element e = S.read();
        send(D[e]);
    }
}
\end{lstlisting}

\begin{lstlisting}[caption=DRH nodel level, label=ls:DRH_node, mathescape=true, float]
compress(Stream S, float Q) {
	Element lastValue = null;
	Element lastDelta = null;
	int counter = 0;
	foreach (s in S) {
	  if (last != null && lastDelta != null) {
	     float delta = 0;
	     if (Q > 1) {
    	    delta = (int)(lastValue $-$ s) / Q;
	     } else {
	        delta = lastValue $-$ s; 
	     }
	     lastValue = s;
	     if (delta == lastDelta) {
	        counter += 1;
	     } else {
	        encoded = huffmanEncode(lastDelta);
	        send((encoded, counter));
	        lastDelta = delta;
	        counter = 0;
	     }
	  } else {
	     lastValue = s;
	     lastDelta = 0;
	  }
	}
}
\end{lstlisting}

\newpage

\begin{acks}
This research has received funding from the European Union’s Horizon 2020 research and innovation programme under grant agreement No 814496.
\end{acks}

\bibliographystyle{ACM-Reference-Format}

\begin{thebibliography}{1}
\bibitem{asghari_internet_2019}
Asghari, P., Rahmani, A.M., Javadi, H.: Internet of Things applications: A systematic review.
\newblock Computer Networks \textbf{148}, 241--261 (2019).
\newblock \doi{10.1016/j.comnet.2018.12.008}.

\bibitem{ronkainen_designing_2015}
Ronkainen, J., Iivari, A.: Designing a {Data} {Management} {Pipeline} for
  {Pervasive} {Sensor} {Communication} {Systems}.
\newblock Procedia Computer Science \textbf{56}, 183--188 (2015).
\newblock \doi{10.1016/j.procs.2015.07.193}.

\bibitem{jensen_time_2017}
Jensen, S.K., Pedersen, T.B., Thomsen, C.: Time {Series} {Management}
  {Systems}: {A} {Survey}.
\newblock IEEE Transactions on Knowledge and Data Engineering \textbf{29}(11),
  2581--2600 (2017).
\newblock \doi{10.1109/TKDE.2017.2740932}.

\bibitem{salomon_data_2007}
Salomon, D.: Data compression: the complete reference, 4th ed edn.
\newblock Springer, London (2007).

\bibitem{wolff_computing_2003}
Wolff, J.: Wolff - 1990 - Simplicity and Power - Some Unifying Ideas in Comp.pdf.
\newblock CoRR \textbf{cs.AI/0307013} (2003).

\bibitem{sayood_introduction_2006}
Sayood, K.: Introduction to data compression, 3rd ed edn.
\newblock Morgan {Kaufmann} series in multimedia information and systems.
  Elsevier, Amsterdam ; Boston (2006)

\bibitem{marascu_tristan_2014}
Marascu, A., Pompey, P., Bouillet, E., Wurst, M., Verscheure, O., Grund, M.,
  Cudre-Mauroux, P.: {TRISTAN}: {Real}-time analytics on massive time series
  using sparse dictionary compression.
\newblock In: 2014 {IEEE} {International} {Conference} on {Big} {Data} ({Big}
  {Data}), pp. 291--300. IEEE, Washington, DC, USA (2014).
\newblock \doi{10.1109/BigData.2014.7004244}.

\bibitem{mairal_online_2009}
Mairal, J., Bach, F., Ponce, J., Sapiro, G.: Online learning for matrix
  factorization and sparse coding.
\newblock Journal of Machine Learning Research \textbf{11}, 19--60 (2009).
\newblock \doi{10.1145/1756006.1756008}.

\bibitem{mallat_matching_1993}
{Mallat}, S., {Zhifeng Zhang}: Matching pursuits with time-frequency
  dictionaries.
\newblock IEEE Transactions on Signal Processing \textbf{41}(12), 3397--3415
  (1993).
\newblock \doi{10.1109/78.258082}.

\bibitem{khelifati_corad_nodate}
Khelifati, A., Khayati, M., Cudre-Mauroux, P.: Corad: Correlation-aware
  compression of massive time series using sparse dictionary coding.
\newblock In: 2019 IEEE International Conference on Big Data (Big Data), pp.
  2289--2298. IEEE Computer Society, Los Alamitos, CA, USA (2019).
\newblock \doi{10.1109/BigData47090.2019.9005580}.

\bibitem{storer_data_1982}
Storer, J.A., Szymanski, T.G.: Data compression via textual substitution.
\newblock Journal of the ACM (JACM) \textbf{29}(4), 928--951 (1982).
\newblock \doi{10.1145/322344.322346}.

\bibitem{pope_accelerometer_2018}
Pope, J., Vafeas, A., Elsts, A., Oikonomou, G., Piechocki, R., Craddock, I.: An
  accelerometer lossless compression algorithm and energy analysis for {IoT}
  devices.
\newblock In: 2018 {IEEE} {Wireless} {Communications} and {Networking}
  {Conference} {Workshops} ({WCNCW}), pp. 396--401. IEEE, Barcelona (2018).
\newblock \doi{10.1109/WCNCW.2018.8368985}.

\bibitem{tuong_ly_le_lossless_2018}
{Le}, T.L., {Vo}, M.: Lossless data compression algorithm to save energy in
  wireless sensor network.
\newblock In: 2018 4th International Conference on Green Technology and
  Sustainable Development (GTSD), pp. 597--600 (2018).
\newblock \doi{10.1109/GTSD.2018.8595614}.

\bibitem{eichinger_time-series_2015}
Eichinger, F., Efros, P., Karnouskos, S., B{\"o}hm, K.: A time-series
  compression technique and its application to the smart grid.
\newblock The VLDB Journal \textbf{24}(2), 193--218 (2015).
\newblock \doi{10.1007/s00778-014-0368-8}.

\bibitem{lazaridis_capturing_2003}
{Lazaridis}, I., {Mehrotra}, S.: Capturing sensor-generated time series with
  quality guarantees.
\newblock In: Proceedings 19th International Conference on Data Engineering
  (Cat. No.03CH37405), pp. 429--440 (2003).
\newblock \doi{10.1109/ICDE.2003.1260811}

\bibitem{dalari_approximations_2006}
{Dalai}, M., {Leonardi}, R.: Approximations of one-dimensional digital signals
  under the$l^infty$norm.
\newblock IEEE Transactions on Signal Processing \textbf{54}(8), 3111--3124
  (2006).
\newblock \doi{10.1109/TSP.2006.875394}

\bibitem{seidel_small_1991}
Seidel, R.: Small-dimensional linear programming and convex hulls made easy.
\newblock Discrete {\&} Computational Geometry \textbf{6}(3), 423--434 (1991).
\newblock \doi{10.1007/BF02574699}.

\bibitem{hawkins_algorithm_2012}
Hawkins, S.E.I., Darlington, E.H.: Algorithm for {Compressing} {Time}-{Series}
  {Data}.
\newblock NASA Tech Briefs  (2012).
\newblock \url{https://ntrs.nasa.gov/search.jsp?R=20120010460}.

\bibitem{lv_chebyshev_2017}
Lv, X., Shen, S.: On {Chebyshev} polynomials and their applications.
\newblock Advances in Difference Equations \textbf{2017}(1), 343 (2017).
\newblock \doi{10.1186/s13662-017-1387-8}.

\bibitem{olkkonen_ecg_2011}
Abo-Zahhad, M.: {ECG} {Signal} {Compression} {Using} {Discrete} {Wavelet}
  {Transform}.
\newblock In: J.T. Olkkonen (ed.) Discrete {Wavelet} {Transforms} - {Theory}
  and {Applications}. InTech (2011).
\newblock \doi{10.5772/16019}.

\bibitem{Goodfellow-et-al-2016}
Goodfellow, I., Bengio, Y., Courville, A.: Deep Learning.
\newblock MIT Press (2016).
\newblock \url{http://www.deeplearningbook.org}

\bibitem{sherstinsky_fundamentals_2020}
Sherstinsky, A.: Fundamentals of {Recurrent} {Neural} {Network} ({RNN}) and
  {Long} {Short}-{Term} {Memory} ({LSTM}) network.
\newblock Physica D: Nonlinear Phenomena \textbf{404}, 132306 (2020).
\newblock \doi{10.1016/j.physd.2019.132306}.

\bibitem{hsu_time_2017}
Hsu, D.: Time {Series} {Compression} {Based} on {Adaptive} {Piecewise}
  {Recurrent} {Autoencoder}.
\newblock ArXiv: 1707.07961.

\bibitem{sepp_long_1997}
Hochreiter, S., Schmidhuber, J.: Long short-term memory.
\newblock Neural computation \textbf{9}, 1735--80 (1997).
\newblock \doi{10.1162/neco.1997.9.8.1735}.
  
\bibitem{sakr_delta_2018}
Suel, T.: Delta {Compression} {Techniques}.
\newblock In: S.~Sakr, A.~Zomaya (eds.) Encyclopedia of {Big} {Data}
  {Technologies}, pp. 1--8. Springer International Publishing, Cham (2018).
\newblock \doi{10.1007/978-3-319-63962-8_63-1}.

\bibitem{hardi_comparative_2019}
Hardi, S.M., Angga, B., Lydia, M.S., Jaya, I., Tarigan, J.T.: Comparative
  {Analysis} {Run}-{Length} {Encoding} {Algorithm} and {Fibonacci} {Code}
  {Algorithm} on {Image} {Compression}.
\newblock Journal of Physics: Conference Series \textbf{1235}, 012107 (2019).
\newblock \doi{10.1088/1742-6596/1235/1/012107}.

\bibitem{walder_ffaasstt_nodate}
Walder, J., Kr{\'{a}}tk{\'{y}}, M., Platos, J.: Fast fibonacci encoding
  algorithm.
\newblock In: J.~Pokorn{\'{y}}, V.~Sn{\'{a}}sel, K.~Richta (eds.) Proceedings
  of the Dateso 2010 Annual International Workshop on DAtabases, TExts,
  Specifications and Objects, Stedronin-Plazy, Czech Republic, April 21-23,
  2010, \emph{{CEUR} Workshop Proceedings}, vol. 567, pp. 72--83. CEUR-WS.org
  (2010).
\newblock \url{http://ceur-ws.org/Vol-567/paper14.pdf}.

\bibitem{mogahed_development_2018}
Mogahed, H.S., Yakunin, A.G.: Development of a {Lossless} {Data} {Compression}
  {Algorithm} for {Multichannel} {Environmental} {Monitoring} {Systems}.
\newblock In: 2018 {XIV} {International} {Scientific}-{Technical} {Conference}
  on {Actual} {Problems} of {Electronics} {Instrument} {Engineering} ({APEIE}),
  pp. 483--486. IEEE, Novosibirsk (2018).
\newblock \doi{10.1109/APEIE.2018.8546121}.

\bibitem{blalock_sprintz_2018}
Blalock, D., Madden, S., Guttag, J.: Sprintz: {Time} {Series} {Compression} for
  the {Internet} of {Things}.
\newblock Proceedings of the ACM on Interactive, Mobile, Wearable and
  Ubiquitous Technologies \textbf{2}(3), 1--23 (2018).
\newblock \doi{10.1145/3264903}.

\bibitem{spiegel_comparative_2018}
Spiegel, J., Wira, P., Hermann, G.: A {Comparative} {Experimental} {Study} of
  {Lossless} {Compression} {Algorithms} for {Enhancing} {Energy} {Efficiency}
  in {Smart} {Meters}.
\newblock In: 2018 {IEEE} 16th {International} {Conference} on {Industrial}
  {Informatics} ({INDIN}), pp. 447--452. IEEE, Porto (2018).
\newblock \doi{10.1109/INDIN.2018.8471921}.

\bibitem{campobello_rake_2017}
Campobello, G., Segreto, A., Zanafi, S., Serrano, S.: {RAKE}: {A} simple and
  efficient lossless compression algorithm for the {Internet} of {Things}.
\newblock In: 2017 25th {European} {Signal} {Processing} {Conference}
  ({EUSIPCO}), pp. 2581--2585. IEEE, Kos, Greece (2017).
\newblock \doi{10.23919/EUSIPCO.2017.8081677}.

\bibitem{fink_compression_2011}
Fink, E., Gandhi, H.S.: Compression of time series by extracting major extrema.
\newblock Journal of Experimental \& Theoretical Artificial Intelligence
  \textbf{23}(2), 255--270 (2011).
\newblock \doi{10.1080/0952813X.2010.505800}.

\bibitem{goldstein_real-time_nodate}
Goldstein, R., Glueck, M., Khan, A.: Real-time compression of time series
  building performance data.
\newblock In: Proceedings of IBPSA-AIRAH Building Simulation Conference (2011).

\bibitem{inamura_keyframe_2003}
Inamura, T., Tanie, H., Nakamura, Y.: Keyframe compression and decompression
  for time series data based on the continuous hidden {Markov} model.
\newblock In: Proceedings 2003 {IEEE}/{RSJ} {International} {Conference} on
  {Intelligent} {Robots} and {Systems} ({IROS} 2003) ({Cat}. {No}.{03CH37453}),
  vol.~2, pp. 1487--1492. IEEE, Las Vegas, NV, USA (2003).
\newblock \doi{10.1109/IROS.2003.1248854}.

\bibitem{young_the_2000}
Steve, Y., Gunnar, E., Mark, G., Thomas, H., Dan, K., Xunying~Andrew, L.,
  Gareth, M., Julian, O., Dave, O., Valtcho, V., Phil, W.: The HTK Book.
\newblock Microsoft (2006)

\bibitem{wojnarski_IEEE_2010}
{Wojnarski}, M., {Gora}, P., {Szczuka}, M., {Nguyen}, H.S., {Swietlicka}, J.,
  {Zeinalipour}, D.: Ieee icdm 2010 contest: Tomtom traffic prediction for
  intelligent gps navigation.
\newblock In: 2010 IEEE International Conference on Data Mining Workshops, pp.
  1372--1376 (2010).
\newblock \doi{10.1109/ICDMW.2010.51}

\bibitem{smith_the_2008}
Smith, K., Depolo, D., Torrisi, J., Edwards, N., Biasi, G., Slater, D.: The
  2008 mw 6.0 wells, nevada earthquake sequence.
\newblock AGU Fall Meeting Abstracts  (2008)

\bibitem{shoaib_fusion_2014}
Shoaib, M., Bosch, S., Incel, O., Scholten, H., Havinga, P.: Fusion of
  smartphone motion sensors for physical activity recognition.
\newblock Sensors (Basel, Switzerland) \textbf{14}, 10146--10176 (2014).
\newblock \doi{10.3390/s140610146}

\bibitem{reiss_towards_2011}
Reiss, A., Stricker, D.: Towards global aerobic activity monitoring.
\newblock In: Proceedings of the 4th International Conference on PErvasive
  Technologies Related to Assistive Environments, PETRA '11. Association for
  Computing Machinery, New York, NY, USA (2011).
\newblock \doi{10.1145/2141622.2141637}.

\bibitem{krawczak_approach_2014}
Krawczak, M. and Szkatuła, G.: An approach to dimensionality reduction in time series.
\newblock In: 2014 Information Sciences, pp.
  15--36 (2014).
\newblock \doi{10.1016/j.ins.2013.10.037}.

\bibitem{ali_concurrent_2021}
Ali, M. and Borgo, R. and Jones, M. W.: Concurrent time-series selections using deep learning and dimension reduction.
\newblock In: 2021 Knowledge-Based Systems, pp.
  107507 (2021).
\newblock \doi{10.1016/j.knosys.2021.107507}.

\bibitem{cooley_algorithm_nodate}
Cooley, J. and Tukey, J. W.: An Algorithm for the Machine Calculation of Complex Fourier Series.

\bibitem{karim_wavelet_2011}
Ali, M. and Borgo, R. and Jones, M. W.: Wavelet {Transform} and {Fast} {Fourier} {Transform} for signal compression: {A} comparative study.
\newblock In: {International} {Conference} on {Electronic} {Devices}, {Systems} and {Applications}, pp.
  280--285 (2011).
\newblock \doi{10.1109/ICEDSA.2011.5959031}.

\bibitem{huang_refined_1999}
Huang, Y. M. and Wu, J. L.: A refined fast 2-d discrete cosine transform algorithm.
\newblock In: IEEE Transactions on Signal Processing, pp.
  904--907 (1999).
\newblock \doi{10.1109/78.747801}.

\bibitem{mess_compression_nodate}
Mess, J. G. and Schmidt, R. and Fey, G. and Dannemann, F.: On the compression of spacecraft housekeeping data using discrete cosine transforms.

\bibitem{provotar_unsupervised_2019}
Provotar, O. I. and Linder, Y. M. and Fey, G. and Veres, M. M.: Unsupervised {Anomaly} {Detection} in {Time} {Series} {Using} {LSTM}-{Based} {Autoencoders}.
\newblock In: {IEEE} {International} {Conference} on {Advanced} {Trends} in {Information} {Theory} 513--517 (2019)
\newblock \doi{10.1109/ATIT49449.2019.9030505}.

\bibitem{goyal_dzip_2021}Goyal, M., Tatwawadi, K., Chandak, S. \& Ochoa, I. DZip: improved general-purpose loss less compression based on novel neural network modeling. {\em 2021 Data Compression Conference (DCC)}. pp. 153-162 (2021,3), https://ieeexplore.ieee.org/document/9418692/

\bibitem{keogh_locally_nodate}Keogh, E., Chakrabarti, K., Mehrotra, S. \& Pazzani, M. Locally Adaptive Dimensionality Reduction for Indexing Large Time Series Databases. 

\bibitem{zhao_time_2019}Zhao, X., Han, X., Su, W. \& Yan, Z. Time series prediction method based on Convolutional Autoencoder and LSTM. {\em 2019 Chinese Automation Congress (CAC)}. pp. 5790-5793 (2019,11), https://ieeexplore.ieee.org/document/8996842/

\bibitem{10.1145/882082.882086}Lin, J., Keogh, E., Lonardi, S. \& Chiu, B. A Symbolic Representation of Time Series, with Implications for Streaming Algorithms. {\em Proceedings Of The 8th ACM SIGMOD Workshop On Research Issues In Data Mining And Knowledge Discovery}. pp. 2-11 (2003), https://doi.org/10.1145/882082.882086

\end{thebibliography}

\end{document}